\documentclass{article}

\usepackage{arxiv}

\usepackage[utf8]{inputenc} 
\usepackage[T1]{fontenc}    
\usepackage{hyperref}       
\usepackage{url}            
\usepackage{booktabs}       
\usepackage{multirow}
\usepackage{amsmath}
\usepackage{amsfonts}       
\usepackage{nicefrac}       
\usepackage{microtype}      
\usepackage{lipsum}		
\usepackage{graphicx}
\usepackage{natbib}
\usepackage{doi}

\title{SCARV: Structure-Constrained Aggregation for Stable Sample Ranking in Redundant NLP Datasets}

\author{ 
	\href{https://orcid.org/0009-0001-9202-7088}{\includegraphics[scale=0.06]{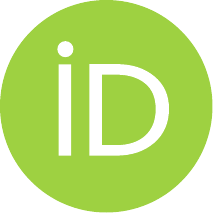}\hspace{1mm}Xu Zheng}
	\hspace{0.75em}
	\href{https://orcid.org/0009-0001-9198-4783}{\includegraphics[scale=0.06]{orcid.pdf}\hspace{1mm}Feiyu Wu}
	\hspace{0.75em}
	\href{https://orcid.org/0009-0001-6477-0480}{\includegraphics[scale=0.06]{orcid.pdf}\hspace{1mm}Linhong Wu}
	\hspace{0.75em}
	\href{https://orcid.org/0009-0009-2355-1191}{\includegraphics[scale=0.06]{orcid.pdf}\hspace{1mm}Zhuocheng Wang}
	\hspace{0.75em}
	\href{https://orcid.org/0000-0001-8310-7169}{\includegraphics[scale=0.06]{orcid.pdf}\hspace{1mm}Hui Li}\thanks{Corresponding author: lihui@mail.xidian.edu.cn} \\
	School of Cyber Engineering, Xidian University \\
	\texttt{zhengxu200477@gmail.com, sn0wm1ans@gmail.com, wlh210121@163.com} \\
	\texttt{smilencet1@gmail.com, lihui@mail.xidian.edu.cn} \\
}

\renewcommand{\shorttitle}{\textit{arXiv} Template}

\hypersetup{
pdftitle={SCARV: Structure-Constrained Aggregation for Stable Sample Ranking in Redundant NLP Datasets},
pdfsubject={Machine Learning},
pdfauthor={Xu Zheng, Feiyu Wu, Linhong Wu, Zhuocheng Wang, Hui Li},
pdfkeywords={sample-level ranking, data-centric NLP, ranking stability, redundant NLP datasets, multi-seed aggregation, structure-aware aggregation},
}

\begin{document}
\maketitle
\begin{abstract}
Sample-level rankings are increasingly used in data-centric NLP for analysis, filtering, debugging, and curation, yet existing pipelines typically score training examples pointwise and rank them as if they were independent. This assumption is fragile in the presence of exact duplicates, near-duplicates, paraphrases, and other redundant structure common in NLP corpora, where stochastic training can make highly similar examples receive unstable relative orderings across random seeds. We study stable sample-level ranking under redundancy and propose \textsc{SCARV}, a modular aggregation framework that operates on top of an existing scoring proxy. \textsc{SCARV} combines robust multi-seed aggregation with a structure-aware aggregation/allocation step over redundancy clusters. Across synthetic redundancy, naturally mined QQP redundancy, multiple proxy families, several NLP tasks, and end-to-end DistilBERT fine-tuning, \textsc{SCARV} substantially improves over bare proxy rankings in global and local stability and yields more reproducible ranking-based decisions such as subset selection and suspicious-example retrieval. Our decomposition and compute-aware frontier sharpen the mechanism: robust multi-seed aggregation is the dominant generic stabilizer, while the structure-aware component adds value mainly under low aggregation budgets or when redundancy clusters are informative, naturally occurring, or sufficiently covered. These results position \textsc{SCARV} not as a universal data selector or a universally dominant replacement for seed-only aggregation, but as a stability-oriented aggregation layer for proxy-induced rankings in redundant NLP datasets.
\end{abstract}

\section{Introduction}

Sample-level ranking has become a core primitive in data-centric NLP. Researchers and practitioners use example-level scores to inspect corpora, identify suspicious or mislabeled training points, filter data, and construct smaller or higher-quality subsets. Existing pipelines derive such rankings from data valuation, training-data attribution, or training-dynamics signals, including Shapley-style valuation, learned valuation, influence-based scoring, example forgetting, and dataset cartography \citep{ghorbani2019data,jia2019towards,yoon2020data,koh2017understanding,pruthi2020estimating,park2023trak,toneva2019empirical,swayamdipta2020dataset}. Although these methods differ in semantics and motivation, they typically share the same operational pattern: assign a pointwise score to each example, then sort the resulting scores as if the examples were independent units.

\begin{figure*}[t]
    \centering
    \includegraphics[width=0.9\textwidth]{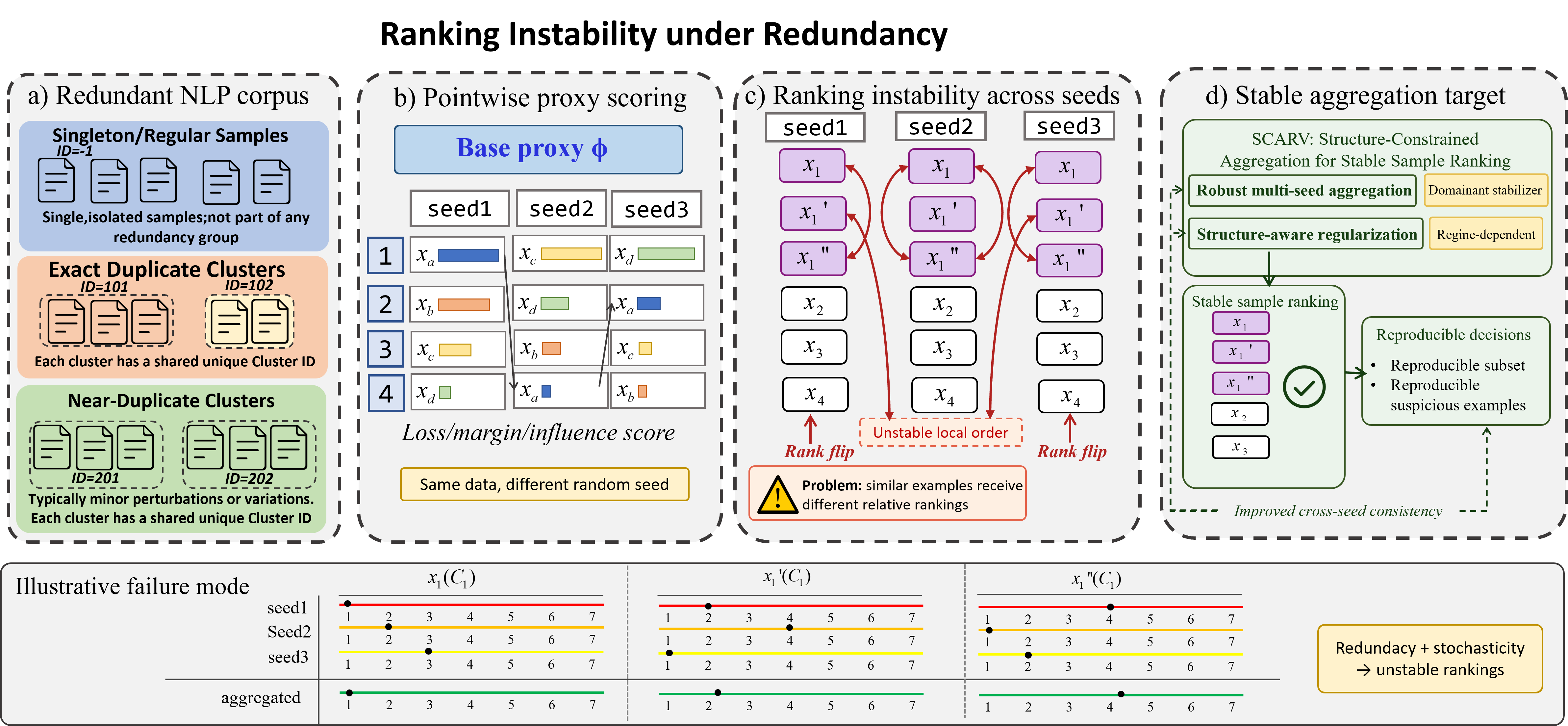}
    \caption{\textbf{Ranking instability under redundancy motivates robust aggregation.} Multi-seed aggregation is the dominant generic stabilization mechanism, while structure-aware aggregation provides regime-dependent additional gains when redundancy clusters are informative.}
    \label{fig:motivation_stability}
\end{figure*}

For NLP corpora, this independence assumption is often fragile. Real datasets contain exact duplicates, near-duplicates, paraphrases, templatic expressions, repeated reports, and other forms of local redundancy. Deduplication studies have shown that repeated text is common enough to affect memorization, privacy risk, and empirical conclusions in modern language-model pipelines \citep{lee2022deduplicating,kandpal2022deduplicating}. Under such redundancy, sample-level ranking becomes unstable in a practically inconvenient way: examples that convey nearly the same information can still receive noticeably different scores because of random initialization, stochastic optimization, bootstrap resampling, or incidental interactions between a proxy and repeated structure. As a result, the local ordering of highly similar examples can vary across random seeds even when the dataset, model family, and base proxy are otherwise fixed. Figure~\ref{fig:motivation_stability} illustrates this failure mode.

This observation suggests that, in redundant NLP datasets, sample ranking is not only a question of \emph{what} signal to use, but also of \emph{how reproducibly} that signal can be realized. Recent work has increasingly recognized robustness as a first-class concern in data valuation: Distributional Shapley emphasizes stability under data perturbations, while Data Banzhaf and Weighted Banzhaf explicitly study how stochastic learning induces inconsistent value rankings \citep{ghorbani2020distributional,wang2023databanzhaf,li2023weightedbanzhaf}. Yet an important practical question remains underexplored for NLP: given an existing sample-level scoring proxy and a dataset with exact or approximate redundancy, how should noisy pointwise scores be aggregated so that the resulting ranking is more stable across seeds?

We study this problem as \emph{stable sample-level ranking under redundancy}. Our goal is deliberately narrower than proposing a universally optimal data selector. We do not seek to replace existing proxies with a new monolithic notion of sample value, nor do we assume that greater ranking stability automatically implies stronger downstream pruning performance. Instead, we ask a modular question: given a base proxy and an approximate redundancy map, how can we transform per-example scores into a ranking that is more stable across random seeds while preserving the main signal of the original proxy? This question is complementary to deduplication. Deduplication asks whether repeated content should be removed or consolidated before training; we ask how rankings should be stabilized once redundancy is already present or approximately identified.

To address this question, we propose \textsc{SCARV} (\textbf{S}tructure-\textbf{C}onstrained \textbf{A}ggregation for \textbf{S}table \textbf{S}ample \textbf{R}anking), a modular aggregation framework that operates on top of an existing proxy. \textsc{SCARV} combines two ingredients: robust multi-seed aggregation, which suppresses seed-specific fluctuations by aggregating proxy scores across repeated runs, and a structure-aware aggregation/allocation step, which uses redundancy clusters to regularize how local scores are pooled and redistributed within groups of similar examples. Conceptually, \textsc{SCARV} does not redefine sample importance; it regularizes how a chosen proxy behaves under redundancy and stochastic training.

A central principle in our study is to separate three objectives that are often conflated in data-centric NLP: \emph{valuation alignment}, \emph{ranking stability}, and \emph{selection utility}. A ranking may correlate well with a gold signal such as leave-one-out yet still be unstable across seeds. Conversely, a stable ranking need not be the strongest selector under a particular retention budget or utility function. This distinction is especially important in light of recent analyses showing that valuation quality and selection quality should not be treated as interchangeable \citep{wang2024rethinking}. We therefore evaluate \textsc{SCARV} primarily through the lens of ranking stability, while treating downstream selection and debugging utility as separate empirical questions rather than automatic consequences of stabilization.

Our empirical results paint a deliberately nuanced picture. \textsc{SCARV} substantially improves over bare proxy rankings across synthetic redundancy, naturally mined QQP redundancy, multiple proxy families, and several NLP tasks. The same instability and stabilization effects also appear under end-to-end DistilBERT fine-tuning, reducing the concern that the phenomenon is an artifact of shallow TF--IDF models. At the same time, our decomposition and compute-aware experiments show that the dominant source of gain is robust multi-seed aggregation rather than the structure-aware module alone. Full \textsc{SCARV} is most competitive at low aggregation budgets and under informative natural redundancy, while seed-mean and seed-Borda remain strong upper baselines for pure stability when enough repeated proxy runs are affordable. Figure~\ref{fig:practical_value} further shows that the practical benefit of stabilization is strongest in the reproducibility of ranking-based decisions, such as subset selection and suspicious-example retrieval, rather than in a universal increase in mean utility.

\paragraph{Contributions.}
First, we formulate stable sample-level ranking under redundancy as a distinct problem in data-centric NLP and argue that valuation alignment, ranking stability, and downstream selection utility should be evaluated separately. Second, we propose \textsc{SCARV}, a modular aggregation framework that combines robust multi-seed aggregation with redundancy-aware structural regularization on top of an existing proxy. Third, we provide a broad empirical study showing that stability-aware aggregation substantially improves bare proxy-induced rankings and decision reproducibility across synthetic redundancy, naturally mined QQP redundancy, multiple proxy families, and several NLP tasks. Fourth, our decomposition, compute-aware, and end-to-end DistilBERT fine-tuning experiments clarify the mechanism and scope of the method: robust multi-seed aggregation is the dominant generic stabilizer, while structure-aware aggregation contributes most in specific regimes, especially when aggregation budgets are small or redundancy clusters are informative, naturally occurring, or sufficiently covered.

\section{Related Work}

\paragraph{Sample-level valuation and scoring.}
Sample-level valuation methods assign importance scores to training examples based on their contribution to model utility. In addition to the Shapley-style and learned valuation methods discussed above, Beta Shapley extends semivalue-based data valuation and broadens the space of admissible value notions \citep{kwon2022beta}. Our work is related in motivation but different in scope: \textsc{SCARV} does not introduce a new valuation functional. Instead, it operates downstream of a chosen proxy and asks how the induced ranking can be made more stable when redundancy and stochastic training interact.

\paragraph{Training-data attribution.}
A closely related line of work studies how individual training examples affect predictions or learned parameters. Beyond influence functions, TracIn, and TRAK, representer-point methods offer another scalable route to example-level attribution \citep{yeh2018representer}. These methods are natural sources of sample-level scores, but they do not by themselves address redundancy-induced ranking instability. \textsc{SCARV} is complementary to this literature: it can sit on top of loss-based or influence-based scores and regularize the final ranking without replacing the underlying attribution mechanism.

\paragraph{Ranking, selection, and data-centric diagnostics.}
Sample-level ranking is also central to neighboring areas such as training-dynamics analysis, data filtering, and subset construction. Alongside example forgetting and dataset cartography, curriculum learning and core-set-style selection study how example order or subset composition affects learning efficiency and downstream performance \citep{bengio2009curriculum,sener2018active}. Our empirical findings are aligned with the broader view that these objectives should not be conflated: \textsc{SCARV} is not a new curriculum or coreset method, and it is not presented as a universal selector. Its primary role is to make proxy-induced rankings more reproducible.

\paragraph{Redundancy, deduplication, and data curation.}
Our setting is especially connected to work on repeated content and deduplication in NLP corpora. Prior studies show that exact and near-exact repetition materially affect memorization, privacy risk, and evaluation reliability in language-model training \citep{lee2022deduplicating,kandpal2022deduplicating}. This literature establishes that redundancy is not a synthetic corner case. \textsc{SCARV} addresses a complementary problem: once duplicate or near-duplicate structure is present or approximately identified, how should sample-level scores be aggregated so that the resulting ranking is more stable across random seeds?

\paragraph{Positioning.}
Taken together, prior work has developed data value notions, attribution mechanisms, and data-centric ranking paradigms for diagnostics or selection. Our paper lies at the intersection of these threads but targets a narrower question: \emph{stable sample-level ranking under redundancy in NLP datasets}. \textsc{SCARV} is therefore best understood as a stability-oriented aggregation layer rather than a new monolithic proxy, a new value functional, or a universal data selector.
\section{Method}

\subsection{Problem Setup}

Let \(D=\{(x_i,y_i)\}_{i=1}^n\) denote a training set. For each random seed
\(r \in \{1,\dots,R\}\), a base sample-scoring proxy \(\phi\) produces one score per
training example,
\begin{equation}
s_i^{(r)} = \phi\big((x_i,y_i); D, \theta^{(r)}\big),
\end{equation}
where \(\theta^{(r)}\) is the model obtained under seed \(r\). The proxy \(\phi\) may
be loss-based, influence-oriented, confidence-based, or any other scoring rule that
returns per-example values.

In addition, we assume access to an approximate redundancy map
\(\mathcal{C}=\{C_1,\dots,C_K\}\) over the training set. Each example belongs to
exactly one cluster, where singleton clusters represent samples for which no
redundant partner is identified. The clusters may come from exact duplicates,
synthetic near-duplicates, or naturally mined similarity structure, and they need
not be perfect.

Our goal is not to redefine the proxy itself. Instead, given the noisy pointwise
scores \(\{s_i^{(r)}\}\) and a redundancy map \(\mathcal{C}\), we seek a final ranking
\(\pi\) that is \emph{more stable across random seeds} while preserving the primary
signal of the underlying proxy.

\subsection{SCARV: Stability-Oriented Aggregation}

\begin{figure*}[t]
    \centering
    \includegraphics[width=0.9\textwidth]{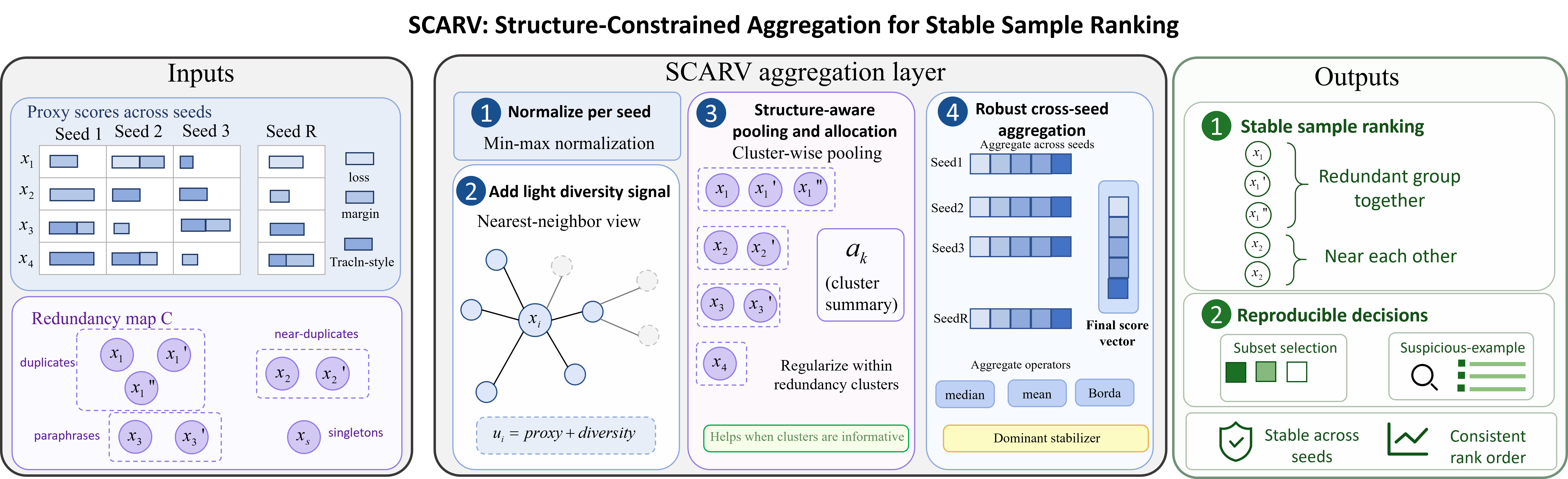}
    \caption{\textbf{SCARV pipeline: proxy scoring, multi-seed aggregation, and structure-aware adjustment.} Robust multi-seed aggregation contributes the largest generic stability gain, while structure-aware allocation acts as a redundancy-dependent regularizer that helps most when clusters are informative.}
    \label{fig:scarv_overview}
\end{figure*}

\textsc{SCARV} (\textbf{S}tructure-\textbf{C}onstrained \textbf{A}ggregation for
\textbf{S}table \textbf{S}ample \textbf{R}anking) is a modular aggregation layer on
top of an existing proxy. Figure~\ref{fig:scarv_overview} summarizes the
pipeline. Starting from per-seed proxy scores, SCARV first constructs a
redundancy-aware score within each seed and then aggregates these scores across
seeds with a robust operator. The framework is designed to stabilize a
proxy-induced ranking rather than to define a new valuation functional. It
regularizes how an existing scoring signal behaves under redundancy and
stochastic training.

\paragraph{Within-seed score preparation and structure-aware aggregation.}
Because raw proxy magnitudes may vary across seeds, we first normalize scores
within each run:
\begin{equation}
\bar{s}_i^{(r)} =
\frac{s_i^{(r)} - \min_j s_j^{(r)}}
     {\max_j s_j^{(r)} - \min_j s_j^{(r)} + \varepsilon}.
\end{equation}
In the default implementation, we then add a light diversity term,
\begin{equation}
u_i^{(r)} = (1-w)\bar{s}_i^{(r)} + w d_i,
\qquad w=0.2,
\end{equation}
where \(d_i \in [0,1]\) is the normalized mean cosine distance to the \(k=10\)
nearest neighbors of example \(i\). This preprocessing keeps the proxy as the
dominant signal while reducing ties among highly similar examples.

Let \(c(i)\in\{1,\dots,K\}\) denote the cluster index of example \(i\). For each
cluster \(C_k\), SCARV computes a cluster-level summary
\begin{equation}
a_k^{(r)} = A\big(\{u_j^{(r)} : j \in C_k\}\big),
\end{equation}
and then allocates this signal back to cluster members through
\begin{equation}
\hat{s}_i^{(r)} =
B\big(u_i^{(r)}, a_{c(i)}^{(r)}, |C_{c(i)}|\big).
\end{equation}
Here \(A(\cdot)\) is a cluster aggregation operator and \(B(\cdot)\) is a
cluster-aware allocation rule. Conceptually, this stage shrinks noisy pointwise
scores toward a shared cluster signal while leaving singleton clusters unchanged.
The redundancy map thus acts as a structural regularizer rather than as a
replacement for the base proxy.

\paragraph{Robust cross-seed aggregation.}
After the within-seed stage, each example has one score per seed,
\(\hat{s}_i^{(1)},\dots,\hat{s}_i^{(R)}\). SCARV then forms the final score
\begin{equation}
\tilde{s}_i =
\mathrm{Agg}\big(\hat{s}_i^{(1)},\dots,\hat{s}_i^{(R)}\big),
\end{equation}
and ranks samples by
\begin{equation}
\pi_{\mathrm{SCARV}} = \operatorname{rank}\big(\tilde{s}_1,\dots,\tilde{s}_n\big).
\end{equation}
In the default instantiation, \(\mathrm{Agg}(\cdot)\) is the median:
\begin{equation}
\tilde{s}_i =
\operatorname{median}\big(\hat{s}_i^{(1)},\dots,\hat{s}_i^{(R)}\big).
\end{equation}
This robust aggregation suppresses seed-specific outliers while preserving the
central tendency of the proxy signal.

Although the default instantiation uses the median for robustness, we explicitly
compare against seed-mean and seed-Borda aggregation in the experiments. This
comparison is important because strong seed-only aggregation can be an upper
baseline for pure stability when enough repeated proxy runs are available.

\paragraph{Special cases used in our experiments.}
The framework subsumes several informative variants that we compare explicitly in
Table~\ref{tab:decomposition}:
\begin{itemize}
    \item \textbf{Bare proxy:} no structure-aware stage and no cross-seed aggregation.
    \item \textbf{Seed-only aggregation:} bypass the structure-aware stage and apply only cross-seed aggregation. We report seed-median as the default robust baseline and also evaluate seed-mean and seed-Borda as stronger pure-stability variants.
    \item \textbf{Cluster-only aggregation:} apply the within-seed structural stage
    but do not aggregate across seeds.
    \item \textbf{Full SCARV:} use both the structure-aware and cross-seed stages.
\end{itemize}
This nested formulation is useful because it separates three possible sources of
improvement: the base proxy itself, robust multi-seed aggregation, and
redundancy-aware structural regularization.

\paragraph{Scope.}
SCARV is proxy-agnostic: any scoring rule \(\phi\) can be plugged in as long as it
produces per-example scores across seeds. In our experiments, we instantiate this
interface with loss-based, influence-style, hybrid, and confidence-based proxies.
Following the distinction introduced in Section~1, we evaluate \emph{valuation
alignment}, \emph{ranking stability}, and \emph{selection utility} separately.
SCARV is designed primarily to improve ranking stability while avoiding large changes to the semantics of an informative base proxy; it does not assume that stabilization alone will optimize downstream selection utility \citep{wang2024rethinking}.

\section{Experiments}

 We evaluate \textsc{SCARV} along five questions:
 (1) how stabilization interacts with the semantics of the base proxy,
 (2) whether it improves ranking stability under redundancy,
 (3) what part of the framework is responsible for the gain,
 (4) when the structure-aware stage is useful relative to seed-only aggregation under compute-aware budgets, and
 (5) whether improved stability makes ranking-based decisions more reproducible.
 
 Following our framing, we treat valuation alignment, ranking stability, and downstream utility as related but distinct targets.

 \subsection{Experimental Setup}

 We use SST-2, MRPC, MNLI, and QQP from GLUE \citep{wang2018glue}, together with AG News \citep{zhang2015character}. Unless otherwise stated, the controlled experiments use TF--IDF features with logistic regression, which allows us to run broad sweeps over redundancy, proxy, clustering, and seed-budget conditions. To address whether the phenomenon is an artifact of shallow models, we additionally run an end-to-end fine-tuning suite with \texttt{distilbert-base-uncased} on SST-2, MRPC, QQP, and MNLI. We study exact duplicates, synthetic near-duplicates, and naturally mined redundancy. The audited default synthetic regime uses redundancy rate \(r=0.3\) and perturbation strength \(s=0.6\).

 Unless noted otherwise, reported main-text numbers use \(R=5\) internal aggregation seeds. For aggregation methods, one outer run produces a final ranking from a group of \(R\) internal scoring seeds, and stability is then computed by comparing final rankings across outer runs rather than by treating seed-pair correlations within the same group as independent observations.

 For the compute-aware frontier, we compare Full \textsc{SCARV} against a post-hoc best seed-only upper baseline under the same seed budget, defined as the maximum over seed-mean, seed-median, and seed-Borda.

 \paragraph{Alignment.}
 We use LOO alignment only as a diagnostic rather than as a primary target. The appendix reports small-scale LOO checks showing a mixed picture: TracIn-style scores are generally more positively aligned with LOO than loss or margin proxies, while stability-aware aggregation can improve, preserve, or reduce LOO alignment depending on the dataset and proxy. We therefore do not treat \textsc{SCARV} as a method for improving valuation alignment. Its role is to stabilize a chosen proxy-induced ranking.

 \subsection{Stable Ranking under Synthetic Redundancy and Across Tasks}

 Table~\ref{tab:main_stability} reports the main stability results. Two patterns are consistent. First, both seed-median aggregation and full \textsc{SCARV} substantially outperform the bare proxy across tasks, confirming that redundancy-aware aggregation addresses a real instability problem rather than a dataset-specific artifact. Second, the relative ordering between seed-median and full \textsc{SCARV} is regime dependent. In the audited SST-2 synthetic setting, the default seed-median baseline is stronger than full \textsc{SCARV} for the loss proxy, whereas on MRPC and MNLI full \textsc{SCARV} is competitive with or slightly above seed-median.

 \begin{table*}[t]
     \centering
     \small
    \caption{\textbf{Global ranking stability across tasks.} Mean pairwise Spearman correlation across random seeds. In the default synthetic SST-2 regime, seed-median is strongest on the loss proxy; on MRPC and MNLI, full \textsc{SCARV} is competitive with or slightly above seed-median.}
     \label{tab:main_stability}
     \setlength{\tabcolsep}{6pt}
     \begin{tabular}{llccc}

        \toprule
        \textbf{Setting} & \textbf{Proxy} & \textbf{Bare} & \textbf{Seed-median} & \textbf{Full SCARV} \\
        \midrule
        SST-2 (synthetic) & Loss & 0.6040 & \textbf{0.8422} & 0.8018 \\
        SST-2 (synthetic) & TracIn-style & 0.9564 & 0.9769 & \textbf{0.9774} \\
        MRPC (TF--IDF) & Loss & 0.8092 & 0.9223 & \textbf{0.9368} \\
        MRPC (TF--IDF) & TracIn-style & 0.8998 & 0.9276 & \textbf{0.9351} \\
        MNLI (TF--IDF) & Loss & 0.3337 & 0.6330 & \textbf{0.6395} \\
        MNLI (TF--IDF) & TracIn-style & 0.8793 & 0.9316 & \textbf{0.9388} \\
        \bottomrule
     \end{tabular}
 \end{table*}

 These results substantially weaken the interpretation that ranking instability is confined to a single toy setting. The same qualitative trend transfers beyond the original SST-2/AG News setup to paraphrase matching (MRPC) and NLI (MNLI). At the same time, Table~\ref{tab:main_stability} already illustrates the central mechanism of the paper: full \textsc{SCARV} is reliably above the bare proxy, but its comparison to seed-median is regime-dependent.
 We therefore treat the method as a stability-oriented aggregation layer rather than as a universally dominant replacement for the seed-median baseline.

 \subsection{Robustness and Mechanism}

 We next test whether the gains are artifacts of the original fixed-slice design and what part of the method is actually responsible for them. Under five stratified subset draws and five redundancy-layout seeds on SST-2 and AG News, mean stability increases from \(0.7556\) for the bare proxy to \(0.8970\) for full \textsc{SCARV} and \(0.9012\) for seed-median. At a selection budget of \(0.3\), subset overlap rises from \(0.5415\) to \(0.6699\) (\textsc{SCARV}) and \(0.6803\) (seed-median), while the selection gap drops from \(0.0309\) to \(0.0232\) and \(0.0239\), respectively. Thus, the stability story is robust to subset-draw and layout variation.

 Table~\ref{tab:decomposition} then decomposes the mechanism on the audited synthetic regime.

 \begin{table}[t]
     \centering
     \small
    \caption{\textbf{Mechanism decomposition on the audited synthetic regime.} Values are mean Spearman stability averaged across proxies and stress regimes. Robust multi-seed aggregation is the dominant source of gain in this regime: seed-mean, seed-Borda, and seed-median are stronger pure-stability baselines. Full \textsc{SCARV} nevertheless remains substantially above bare, cluster-only, and dedup-first baselines, supporting its role as a structural regularizer rather than a universally dominant component.}
     \label{tab:decomposition}
     \begin{tabular}{@{}lc@{}}

        \toprule
        \textbf{Method} & \textbf{Mean Spearman} \\
        \midrule
        Approx.\ dedup-then-rank & 0.7350 \\
        Cluster-only & 0.7861 \\
        Bare & 0.7880 \\
        Full SCARV & 0.8899 \\
        Seed-median & 0.9159 \\
        Seed-Borda & 0.9445 \\
        Seed-mean & \textbf{0.9470} \\
        \bottomrule
     \end{tabular}
 \end{table}

 The most important empirical fact in Table~\ref{tab:decomposition} is that robust multi-seed aggregation is the dominant source of stability gain in the default synthetic setting: seed-mean, seed-Borda, and seed-median all outperform full \textsc{SCARV} on average. This rules out a strong interpretation in which the structure-aware stage is the universally dominant component. At the same time, full \textsc{SCARV} remains well above bare, cluster-only, and dedup-first baselines. We therefore interpret the structure-aware stage as a regime-dependent structural regularizer that can be useful when redundancy structure is informative, while repeated seed aggregation remains the strongest generic stabilizer.
 
  This regime dependence is also reflected in the paired comparisons. On SST-2 synthetic near-duplicates, full \textsc{SCARV} trails seed-median (\(\Delta=-0.0121\), 95\% CI \([-0.0178,-0.0060]\), \(p=0.0004\)), whereas on QQP natural redundancy the sign reverses in favor of full \textsc{SCARV} (\(\Delta=0.0140\), 95\% CI \([0.0127,0.0154]\), \(p=0.0312\)). MRPC and MNLI show small positive differences for full \textsc{SCARV} but should be interpreted cautiously under the current outer-run sample size. This is precisely why we frame the method as a stability-oriented aggregation layer rather than as a claim of universal structural dominance.

  \subsection{Natural Redundancy}

  A central concern for redundancy-based methods is whether they work only under synthetic perturbations. Table~\ref{tab:natural_redundancy} addresses this issue on QQP. Compared with the earlier weak SST-2 natural setting, naturally mined QQP clusters provide a more informative test bed: sentence-embedding clustering covers \(6.5\%\) of the training set, with 74 non-singleton clusters, mean cluster size \(2.65\), and maximum size \(8\). Although this coverage is still modest, it is non-trivial and sufficient for testing whether the structure-aware stage helps under naturally mined redundancy.

  \begin{table}[t]
      \centering
      \scriptsize
     \caption{\textbf{Natural redundancy on QQP.} Coverage is the fraction of training examples in non-singleton clusters. Values compare the seed-median baseline with Full \textsc{SCARV}. Stronger natural redundancy makes the structure-aware stage useful relative to the default robust-median baseline.}
     \label{tab:natural_redundancy}
     \setlength{\tabcolsep}{4pt}
     \begin{tabular}{@{}lccc@{}}

        \toprule
        \textbf{Cluster mining} & \textbf{Coverage} & \textbf{Loss} & \textbf{TracIn} \\
         & \textbf{(\%)} & \textbf{Seed-med. / Full} & \textbf{Seed-med. / Full} \\
        \midrule
        TF--IDF cosine & 2.6 & 0.936 / 0.948 & 0.907 / 0.922 \\
        Sentence-embedding & 6.5 & 0.936 / 0.948 & 0.907 / 0.923 \\
        Lexical char-gram & 6.1 & 0.936 / 0.948 & 0.907 / 0.924 \\
        \bottomrule
     \end{tabular}
 \end{table}

 In this regime, full \textsc{SCARV} can exceed the default seed-median baseline across the three cluster-discovery procedures reported here and both proxy families. This natural-redundancy result matters for two reasons. First, it shows that the paper is not restricted to synthetic duplicate injection. Second, it clarifies when the structure-aware stage is most useful: redundancy-aware aggregation is most helpful when redundancy is more informative, naturally mined, or sufficiently covered. As the compute-aware frontier below shows, this should not be read as a universal win over all seed-only aggregators; seed-mean and seed-Borda remain strong baselines when enough repeated proxy runs are available.

 \subsection{Compute-Aware and Transformer Robustness}

 A natural objection is that \textsc{SCARV} may be unnecessary once stronger seed-only baselines are allowed, or that the observed instability may be an artifact of shallow TF--IDF models. We address both concerns with two additional audited experiments.

 \begin{table}[t]
    \centering
    \small
    \caption{\textbf{Compute-aware frontier.} $\Delta_{\mathrm{best}}$ is Full \textsc{SCARV} Spearman minus the post-hoc strongest seed-only method among seed-mean, seed-median, and seed-Borda under the same seed budget. This column should be read as an upper baseline for pure stability, not as a fixed deployable method. Full \textsc{SCARV} is strongest mainly at very small budgets; with enough seeds, seed-mean or seed-Borda usually becomes the strongest pure-stability baseline.}
    \label{tab:compute-frontier-main}
    \setlength{\tabcolsep}{3.5pt}
    \begin{tabular}{rrrrr}
        \toprule
        $R$ & \textbf{Settings} & \textbf{Full $>$ best upper} & \textbf{Mean $\Delta_{\mathrm{best}}$} & \textbf{Median $\Delta_{\mathrm{best}}$} \\
        \midrule
        1  & 18 & 16 & +0.0180 & +0.0233 \\
        2  & 18 & 15 & +0.0095 & +0.0126 \\
        3  & 18 & 0 & -0.0374 & -0.0335 \\
        5  & 18 & 0 & -0.0374 & -0.0307 \\
        7  & 18 & 0 & -0.0318 & -0.0251 \\
        10 & 18 & 0 & -0.0187 & -0.0138 \\
        \bottomrule
    \end{tabular}
\end{table}

  First, we run a compute-aware seed-budget frontier with \(R \in \{1,2,3,5,7,10\}\), comparing full \textsc{SCARV} against a post-hoc best seed-only upper baseline under the same aggregation budget. This upper baseline is the maximum over seed-mean, seed-median, and seed-Borda, and should be read as a conservative reference for pure stability rather than as a fixed deployment rule. The result sharpens the mechanism rather than overturning it. At very small budgets, full \textsc{SCARV} is often competitive or stronger: for \(R=1\) it exceeds the best seed-only upper baseline in 16 of 18 settings, and for \(R=2\) in 15 of 18 settings. Once \(R \ge 3\), however, the strongest seed-only method, usually seed-mean or seed-Borda, overtakes full \textsc{SCARV} in all settings in this frontier. Thus, the structure-aware stage is best interpreted as a low-budget and regime-dependent regularizer, not as a universally stronger substitute for repeated seed-only aggregation.

 \begin{table*}[t]
    \centering
    \small
    \caption{\textbf{End-to-end DistilBERT fine-tuning.} Values are mean Spearman stability averaged over confidence, loss, and margin proxies. Full \textsc{SCARV} substantially improves over bare rankings and is competitive with seed-median, while the post-hoc best upper baseline (the strongest seed-only method, usually seed-mean or seed-Borda) remains strongest for pure stability.}
    \label{tab:transformer-main}
    \setlength{\tabcolsep}{8pt}
    \begin{tabular}{lrrrr}
        \toprule
        \textbf{Dataset} & \textbf{Bare} & \textbf{Seed-med.} & \textbf{Full SCARV} & \textbf{Best upper} \\
        \midrule
        MNLI  & 0.872 & 0.937 & 0.935 & 0.970 \\
        MRPC  & 0.923 & 0.959 & 0.960 & 0.983 \\
        QQP   & 0.924 & 0.959 & 0.959 & 0.981 \\
        SST-2 & 0.853 & 0.921 & 0.919 & 0.965 \\
        \midrule
        Avg.  & 0.893 & 0.944 & 0.943 & 0.975 \\
        \bottomrule
    \end{tabular}
\end{table*}

  Second, we run end-to-end \texttt{distilbert-base-uncased} fine-tuning on SST-2, MRPC, QQP, and MNLI. The transformer results confirm that the instability and stabilization effects are not artifacts of TF--IDF/logistic regression. Averaged over confidence, loss, and margin proxies, full \textsc{SCARV} improves Spearman stability from 0.893 for bare rankings to 0.943, and it also improves local Jaccard overlap and selected-subset reproducibility. Full \textsc{SCARV} is essentially tied with seed-median on average. However, seed-mean and seed-Borda remain stronger pure-stability baselines, reaching about 0.975 average Spearman in the best seed-only upper-baseline column. These results support our main interpretation: \textsc{SCARV} is a useful stability layer over bare proxies and a competitive robust-median-style aggregator, while the post-hoc best seed-only upper baseline remains a strong reference for pure stability when enough seeds are affordable.

 \subsection{Practical Value: Reproducible Ranking-Based Decisions}

\begin{figure*}[t]
      \centering
      \includegraphics[width=0.88\textwidth]{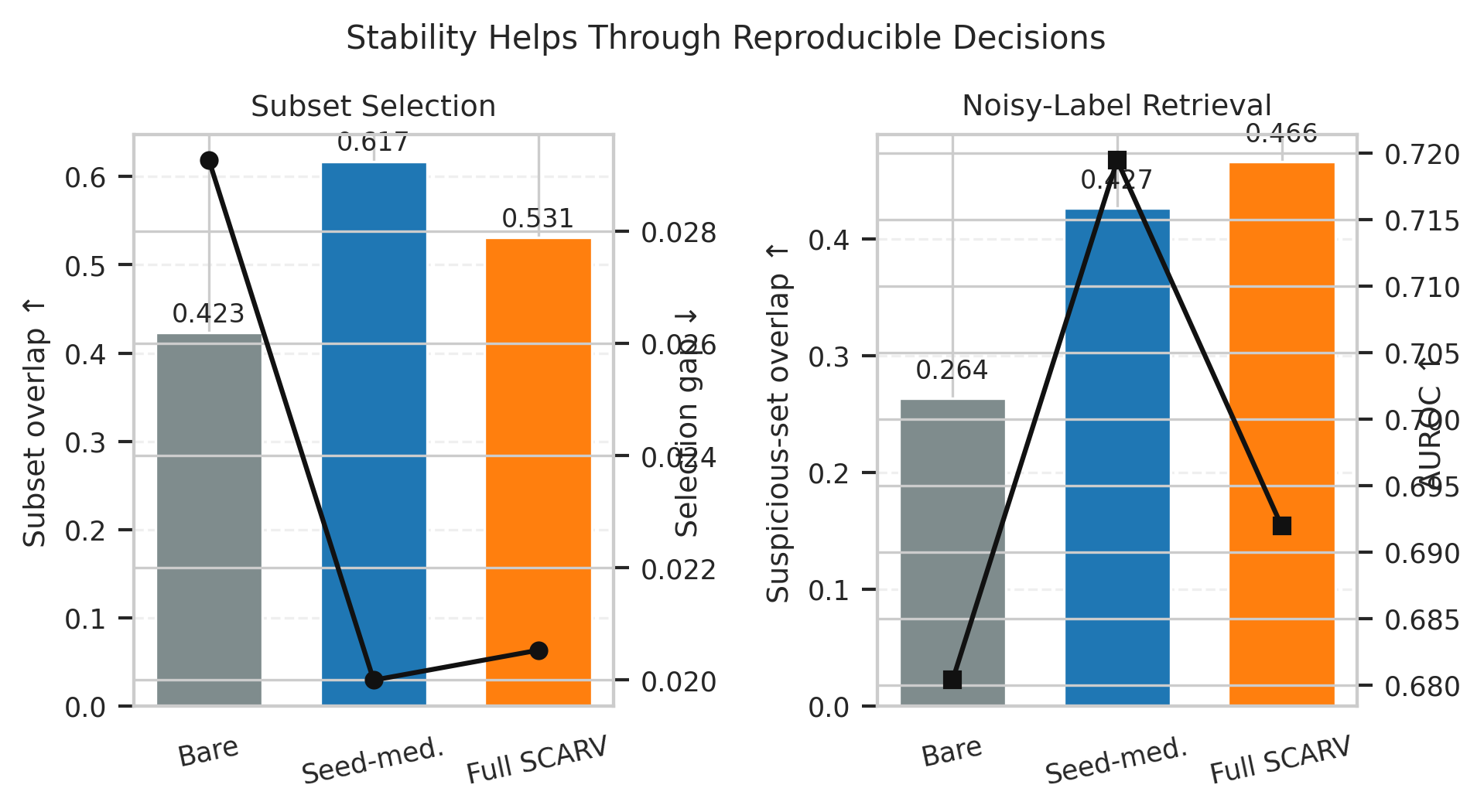}
     \caption{\textbf{Stability has practical value through reproducibility.} Left: under subset selection, stability-aware aggregation substantially increases cross-run overlap of selected subsets and reduces selection gap, even when mean utility changes only modestly. Right: under noisy-label retrieval, seed-median gives the best AUROC in this setting, while full \textsc{SCARV} gives the most reproducible suspicious-example sets.}
      \label{fig:practical_value}
  \end{figure*}

 The strongest practical benefit of stabilization is \emph{decision reproducibility}, not a universal increase in mean utility. Figure~\ref{fig:practical_value} summarizes this result. Under subset selection, seed-median and full \textsc{SCARV} substantially increase overlap between selected subsets across runs and reduce the selection gap, even when mean utility changes only modestly. In the audited SST-2 setting with the loss proxy at budget \(0.3\), subset overlap rises from \(0.4229\) for bare to \(0.6168\) for seed-median and \(0.5306\) for full \textsc{SCARV}, while the selection gap falls from \(0.0293\) to \(0.0200\) and \(0.0205\), respectively.

  The same conclusion holds for noisy-label retrieval. For the loss proxy, seed-median gives the best AUROC in the audited SST-2 setting, improving from \(0.6804\) for bare to \(0.7194\), while full \textsc{SCARV} reaches \(0.6919\). Full \textsc{SCARV} does not dominate AUROC, but it produces the most reproducible suspicious-example sets: the overlap of flagged examples rises from \(0.2636\) for bare to \(0.4270\) for seed-median and \(0.4664\) for full \textsc{SCARV}. Thus, the applied value of \textsc{SCARV} is that it makes ranking-based decisions more repeatable and easier to audit, not that it universally improves every downstream selection metric.

  Overall, the experiments support five conclusions. First, stability-aware aggregation substantially improves over bare proxy-induced rankings under redundancy. Second, the dominant mechanism in the default synthetic regime is robust multi-seed aggregation rather than structure alone. Third, the structure-aware stage is most useful in natural or otherwise informative redundancy regimes and at small aggregation budgets. Fourth, the transformer results reduce the concern that the phenomenon is a shallow-model artifact without overturning the basic mechanism. Fifth, the strongest practical benefit of stabilization is more reproducible ranking-based decisions, while mean selection utility remains proxy-dependent and does not support framing \textsc{SCARV} as a universal selector.
\section{Discussion and Limitations}

\subsection{A deliberately narrow interpretation}
Our results support a deliberately narrow interpretation of \textsc{SCARV}. The method is most useful as a \emph{stability-oriented aggregation layer} for sample-level ranking under redundancy, rather than as a new proxy or a universal selector. Across the main synthetic, realistic-task, natural-redundancy, compute-aware, and transformer settings, the clearest and most consistent effect is improved ranking stability relative to bare proxy rankings. By contrast, downstream subset utility remains strongly proxy- and budget-dependent. This distinction is central to our empirical picture: valuation alignment, ranking stability, and selection utility are related objectives, but they are not interchangeable. The compute-aware and transformer fine-tuning experiments strengthen this interpretation: they show that the phenomenon is not confined to a shallow-model toy setting, while also confirming that best seed-only aggregation remains a strong upper baseline for pure stability.

\subsection{What actually drives the gain}
Our decomposition and compute-aware results clarify the mechanism of the method. In the default synthetic regime, the dominant source of improvement is robust multi-seed aggregation rather than the structure-aware stage alone: seed-only baselines, especially seed-mean and seed-Borda, are often stronger than full \textsc{SCARV} when enough repeated proxy runs are available. The compute frontier makes this boundary explicit. Full \textsc{SCARV} is most competitive at small aggregation budgets, where structural regularization can compensate for limited repeated runs; as the seed budget increases, the best seed-only aggregators usually become the strongest pure-stability baselines. At the same time, full \textsc{SCARV} remains substantially better than bare, cluster-only, and dedup-first baselines, and the natural-redundancy experiments show that the structure-aware component becomes more useful when redundancy is more informative, naturally mined, or sufficiently covered. We therefore view structure-aware aggregation as a regime-dependent structural regularizer rather than a universally dominant component.

\subsection{Why stability matters in practice}
Figure~\ref{fig:practical_value} shows that the main practical value of stabilization is \emph{decision reproducibility}. Stability-aware aggregation makes ranking-based decisions more consistent across runs: selected subsets overlap more, and suspicious-example retrieval surfaces more similar sets of flagged examples across seeds. This benefit is practically meaningful even when mean downstream utility changes only modestly or can move in either direction. In other words, the strongest applied argument for \textsc{SCARV} is not that it universally improves average pruning accuracy, but that it makes proxy-induced decisions more repeatable and therefore easier to trust, compare, and analyze.

\subsection{Implications for modular data-centric pipelines}
These results also support a modular view of data-centric NLP. \textsc{SCARV} does not attempt to replace the semantics of the underlying proxy. Instead, it regularizes how a chosen proxy behaves under redundancy and stochastic training. This perspective is useful precisely because different proxies capture different notions of importance. A proxy that aligns better with a gold valuation criterion need not be the strongest selector under aggressive pruning, and a selector that works well at one retention budget need not induce a stable ranking. Rather than searching for a single universally optimal ranking, it may therefore be more productive to choose a proxy for the desired downstream purpose and then apply a stability-oriented aggregation layer to make its ranking more reproducible.

\subsection{Limitations}
Our study has several limitations. First, our strongest evidence concerns ranking stability and decision reproducibility rather than universal gains in downstream selection utility, so \textsc{SCARV} should not be interpreted as solving the general problem of data subset selection. Second, although the transformer fine-tuning experiments reduce the concern that the phenomenon is a TF--IDF/logistic-regression artifact, the paper still stops short of large-scale pretraining, instruction tuning, or deployment-scale data curation. Third, the structure-aware component depends on the quality and interpretation of the redundancy map. Our structure-quality diagnostics show that pure ranking stability can remain high even under random or degraded cluster assignments, so stability alone should not be taken as evidence that the clusters are semantically meaningful. High ranking stability under structural smoothing should not be interpreted as evidence that the redundancy map is semantically correct; cluster quality should be audited separately. \textsc{SCARV} should therefore be used with cluster-quality auditing rather than arbitrary partitions. Another limitation is that the best seed-only comparison in the compute frontier is a post-hoc upper baseline over several aggregation rules. This makes it useful for stress-testing pure stability claims, but it should not be interpreted as a single fixed baseline that would be chosen without validation in deployment. Finally, the current framework uses hard clusters and relatively simple allocation rules. Future work should study soft or uncertain redundancy maps, learned cluster confidence, and adaptive shrinkage rules that respond to cluster size, similarity, and proxy family.

\bibliographystyle{unsrtnat}
\bibliography{references}  

@inproceedings{ghorbani2019data,
  title     = {Data Shapley: Equitable Valuation of Data for Machine Learning},
  author    = {Ghorbani, Amirata and Zou, James},
  booktitle = {Proceedings of the 36th International Conference on Machine Learning},
  series    = {Proceedings of Machine Learning Research},
  volume    = {97},
  pages     = {2242--2251},
  year      = {2019}
}

@inproceedings{jia2019towards,
  title     = {Towards Efficient Data Valuation Based on the Shapley Value},
  author    = {Jia, Ruoxi and Dao, David and Wang, Boxin and Hubis, Frances Ann and Hynes, Nick and G{\"u}rel, Nezihe Merve and Li, Bo and Zhang, Ce and Song, Dawn and Spanos, Costas J.},
  booktitle = {Proceedings of the 22nd International Conference on Artificial Intelligence and Statistics},
  series    = {Proceedings of Machine Learning Research},
  volume    = {89},
  pages     = {1167--1176},
  year      = {2019}
}

@inproceedings{yoon2020data,
  title     = {Data Valuation Using Reinforcement Learning},
  author    = {Yoon, Jinsung and Arik, Sercan and Pfister, Tomas},
  booktitle = {Proceedings of the 37th International Conference on Machine Learning},
  series    = {Proceedings of Machine Learning Research},
  volume    = {119},
  pages     = {10842--10851},
  year      = {2020}
}

@inproceedings{koh2017understanding,
  title     = {Understanding Black-box Predictions via Influence Functions},
  author    = {Koh, Pang Wei and Liang, Percy},
  booktitle = {Proceedings of the 34th International Conference on Machine Learning},
  series    = {Proceedings of Machine Learning Research},
  volume    = {70},
  pages     = {1885--1894},
  year      = {2017}
}

@inproceedings{pruthi2020estimating,
  title     = {Estimating Training Data Influence by Tracing Gradient Descent},
  author    = {Pruthi, Garima and Liu, Frederick and Kale, Satyen and Sundararajan, Mukund},
  booktitle = {Advances in Neural Information Processing Systems},
  volume    = {33},
  pages     = {19920--19930},
  year      = {2020}
}

@inproceedings{park2023trak,
  title     = {{TRAK}: Attributing Model Behavior at Scale},
  author    = {Park, Sung Min and Georgiev, Kristian and Ilyas, Andrew and Leclerc, Guillaume and Madry, Aleksander},
  booktitle = {Proceedings of the 40th International Conference on Machine Learning},
  series    = {Proceedings of Machine Learning Research},
  volume    = {202},
  pages     = {27074--27113},
  year      = {2023}
}

@inproceedings{toneva2019empirical,
  title     = {An Empirical Study of Example Forgetting during Deep Neural Network Learning},
  author    = {Toneva, Mariya and Sordoni, Alessandro and Tachet des Combes, R{\'e}mi and Trischler, Adam and Bengio, Yoshua and Gordon, Geoffrey J.},
  booktitle = {International Conference on Learning Representations},
  year      = {2019}
}

@inproceedings{swayamdipta2020dataset,
  title     = {Dataset Cartography: Mapping and Diagnosing Datasets with Training Dynamics},
  author    = {Swayamdipta, Swabha and Schwartz, Roy and Lourie, Nicholas and Wang, Yizhong and Hajishirzi, Hannaneh and Smith, Noah A. and Choi, Yejin},
  booktitle = {Proceedings of the 2020 Conference on Empirical Methods in Natural Language Processing (EMNLP)},
  pages     = {9275--9293},
  year      = {2020}
}

@inproceedings{lee2022deduplicating,
  title     = {Deduplicating Training Data Makes Language Models Better},
  author    = {Lee, Katherine and Ippolito, Daphne and Nystrom, Andrew and Zhang, Chiyuan and Eck, Douglas and Callison-Burch, Chris and Carlini, Nicholas},
  booktitle = {Proceedings of the 60th Annual Meeting of the Association for Computational Linguistics (Volume 1: Long Papers)},
  pages     = {8424--8445},
  year      = {2022}
}

@inproceedings{kandpal2022deduplicating,
  title     = {Deduplicating Training Data Mitigates Privacy Risks in Language Models},
  author    = {Kandpal, Nikhil and Wallace, Eric and Raffel, Colin},
  booktitle = {Proceedings of the 39th International Conference on Machine Learning},
  series    = {Proceedings of Machine Learning Research},
  volume    = {162},
  pages     = {10697--10707},
  year      = {2022}
}

@inproceedings{ghorbani2020distributional,
  title     = {A Distributional Framework For Data Valuation},
  author    = {Ghorbani, Amirata and Kim, Michael and Zou, James},
  booktitle = {Proceedings of the 37th International Conference on Machine Learning},
  series    = {Proceedings of Machine Learning Research},
  volume    = {119},
  pages     = {3535--3544},
  year      = {2020}
}

@inproceedings{wang2023databanzhaf,
  title     = {Data Banzhaf: A Robust Data Valuation Framework for Machine Learning},
  author    = {Wang, Jiachen T. and Jia, Ruoxi},
  booktitle = {Proceedings of the 26th International Conference on Artificial Intelligence and Statistics},
  series    = {Proceedings of Machine Learning Research},
  volume    = {206},
  pages     = {6388--6421},
  year      = {2023}
}

@inproceedings{li2023weightedbanzhaf,
  title     = {Robust Data Valuation with Weighted Banzhaf Values},
  author    = {Li, Weida and Yu, Yaoliang},
  booktitle = {Advances in Neural Information Processing Systems},
  volume    = {36},
  year      = {2023}
}

@inproceedings{wang2024rethinking,
  title     = {Rethinking Data Shapley for Data Selection Tasks: Misleads and Merits},
  author    = {Wang, Jiachen T. and Yang, Tianji and Zou, James and Kwon, Yongchan and Jia, Ruoxi},
  booktitle = {Proceedings of the 41st International Conference on Machine Learning},
  series    = {Proceedings of Machine Learning Research},
  volume    = {235},
  pages     = {52033--52063},
  year      = {2024}
}

@inproceedings{kwon2022beta,
  title     = {Beta Shapley: A Unified and Noise-reduced Data Valuation Framework for Machine Learning},
  author    = {Kwon, Yongchan and Zou, James},
  booktitle = {Proceedings of the 25th International Conference on Artificial Intelligence and Statistics},
  series    = {Proceedings of Machine Learning Research},
  volume    = {151},
  pages     = {8780--8802},
  year      = {2022}
}

@inproceedings{yeh2018representer,
  title     = {Representer Point Selection for Explaining Deep Neural Networks},
  author    = {Yeh, Chih-Kuan and Kim, Joon and Yen, Ian En-Hsu and Ravikumar, Pradeep K.},
  booktitle = {Advances in Neural Information Processing Systems},
  volume    = {31},
  year      = {2018}
}

@inproceedings{bengio2009curriculum,
  title     = {Curriculum Learning},
  author    = {Bengio, Yoshua and Louradour, J{\'e}r{\^o}me and Collobert, Ronan and Weston, Jason},
  booktitle = {Proceedings of the 26th Annual International Conference on Machine Learning},
  pages     = {41--48},
  year      = {2009}
}

@inproceedings{sener2018active,
  title     = {Active Learning for Convolutional Neural Networks: A Core-Set Approach},
  author    = {Sener, Ozan and Savarese, Silvio},
  booktitle = {International Conference on Learning Representations},
  year      = {2018}
}

@inproceedings{wang2018glue,
  title     = {{GLUE}: A Multi-Task Benchmark and Analysis Platform for Natural Language Understanding},
  author    = {Wang, Alex and Singh, Amanpreet and Michael, Julian and Hill, Felix and Levy, Omer and Bowman, Samuel R.},
  booktitle = {Proceedings of the 2018 EMNLP Workshop BlackboxNLP: Analyzing and Interpreting Neural Networks for NLP},
  pages     = {353--355},
  year      = {2018}
}

@inproceedings{zhang2015character,
  title     = {Character-level Convolutional Networks for Text Classification},
  author    = {Zhang, Xiang and Zhao, Junbo and LeCun, Yann},
  booktitle = {Advances in Neural Information Processing Systems},
  volume    = {28},
  year      = {2015}
}

@article{jia2019efficient_vldb,
  title   = {Efficient Task-Specific Data Valuation for Nearest Neighbor Algorithms},
  author  = {Jia, Ruoxi and Dao, David and Wang, Boxin and Hubis, Frances Ann and G{\"u}rel, Nezihe Merve and Li, Bo and Zhang, Ce and Spanos, Costas J. and Song, Dawn},
  journal = {Proceedings of the VLDB Endowment},
  volume  = {12},
  number  = {11},
  pages   = {1610--1623},
  year    = {2019}
}

@inproceedings{wang2024weightedknn,
  title     = {Efficient Data Shapley for Weighted Nearest Neighbor Algorithms},
  author    = {Wang, Jiachen T. and Mittal, Prateek and Jia, Ruoxi},
  booktitle = {Proceedings of The 27th International Conference on Artificial Intelligence and Statistics},
  series    = {Proceedings of Machine Learning Research},
  volume    = {238},
  pages     = {2557--2565},
  year      = {2024}
}

@article{wang2023thresholdknn,
  title   = {Threshold {KNN}-Shapley: A Linear-Time and Privacy-Friendly Approach to Data Valuation},
  author  = {Wang, Jiachen T. and Zhu, Yuqing and Wang, Yu-Xiang and Jia, Ruoxi and Mittal, Prateek},
  journal = {arXiv preprint arXiv:2308.15709},
  year    = {2023}
}

@inproceedings{schoch2022csshapley,
  title     = {{CS}-Shapley: Class-Wise Shapley Values for Data Valuation in Classification},
  author    = {Schoch, Stephanie and Xu, Haifeng and Ji, Yangfeng},
  booktitle = {Advances in Neural Information Processing Systems},
  volume    = {35},
  year      = {2022}
}

@inproceedings{jaramillo2021shapleyconformal,
  title     = {Shapley-Value Based Inductive Conformal Prediction},
  author    = {Jaramillo, William Lopez and Smirnov, Evgueni},
  booktitle = {Proceedings of the Tenth Symposium on Conformal and Probabilistic Prediction and Applications},
  series    = {Proceedings of Machine Learning Research},
  volume    = {152},
  pages     = {52--71},
  year      = {2021}
}

@inproceedings{kwon2024datainf,
  title     = {{DataInf}: Efficiently Estimating Data Influence in {LoRA}-Tuned {LLM}s and Diffusion Models},
  author    = {Kwon, Yongchan and Wu, Eric and Wu, Kevin and Zou, James},
  booktitle = {International Conference on Learning Representations},
  year      = {2024}
}

@inproceedings{bae2024approxunroll,
  title     = {Training Data Attribution via Approximate Unrolling},
  author    = {Bae, Juhan and Lin, Wu and Lorraine, Jonathan and Grosse, Roger},
  booktitle = {Advances in Neural Information Processing Systems},
  volume    = {37},
  year      = {2024}
}

@inproceedings{quintas2024multiply,
  title     = {Multiply-Robust Causal Change Attribution},
  author    = {Quintas-Martinez, Victor and Bahadori, Mohammad Taha and Santiago, Eduardo and Mu, Jeff and Heckerman, David},
  booktitle = {Proceedings of the 41st International Conference on Machine Learning},
  series    = {Proceedings of Machine Learning Research},
  volume    = {235},
  pages     = {41821--41840},
  year      = {2024}
}

@inproceedings{deng2025versatile,
  title     = {A Versatile Influence Function for Data Attribution with Non-Decomposable Loss},
  author    = {Deng, Junwei and Tang, Weijing and Ma, Jiaqi W.},
  booktitle = {Proceedings of the 42nd International Conference on Machine Learning},
  series    = {Proceedings of Machine Learning Research},
  volume    = {267},
  pages     = {13256--13276},
  year      = {2025}
}

@inproceedings{basile2023zeroshot,
  title     = {Zero-Shot Data Maps: Efficient Dataset Cartography without Model Training},
  author    = {Basile, Angelo and Franco-Salvador, Marc and Rosso, Paolo},
  booktitle = {Findings of the Association for Computational Linguistics: EMNLP 2023},
  pages     = {8264--8277},
  year      = {2023}
}

@inproceedings{jagielski2023measuring,
  title     = {Measuring Forgetting of Memorized Training Examples},
  author    = {Jagielski, Matthew and Thakkar, Om and Tram{\`e}r, Florian and Ippolito, Daphne and Lee, Katherine and Carlini, Nicholas and Wallace, Eric and Song, Shuang and Thakurta, Abhradeep and Papernot, Nicolas and Zhang, Chiyuan},
  booktitle = {International Conference on Learning Representations},
  year      = {2023}
}

@inproceedings{leybzon2024learning,
  title     = {Learning, Forgetting, Remembering: Insights from Tracking {LLM} Memorization during Training},
  author    = {Leybzon, Danny D. and Kervadec, Corentin},
  booktitle = {Proceedings of the 7th BlackboxNLP Workshop: Analyzing and Interpreting Neural Networks for NLP},
  pages     = {43--57},
  year      = {2024}
}

@inproceedings{meeus2024copyright,
  title     = {Copyright Traps for Large Language Models},
  author    = {Meeus, Matthieu and Shilov, Igor and Faysse, Manuel and de Montjoye, Yves-Alexandre},
  booktitle = {Proceedings of the 41st International Conference on Machine Learning},
  series    = {Proceedings of Machine Learning Research},
  volume    = {235},
  pages     = {35296--35309},
  year      = {2024}
}

\section{Additional Experimental Details}
\label{app:setup}

\subsection{Audit Status and Reporting Policy}

All appendix results in the revised paper come from the \emph{audited rerun pipeline}. Unless explicitly marked as legacy reproduction, we do not use legacy packaged summaries as primary evidence. This matters because the initial draft mixed together feasibility-style results, fixed-layout synthetic injections, and older summary packaging.

The revised paper uses the following reporting policy. First, the default synthetic near-duplicate operating point is redundancy rate \(r=0.3\) and perturbation strength \(s=0.6\). Earlier references to \(s=0.3\) correspond to an outdated feasibility configuration and are not used as the main support for the revised claims. Second, legacy sequential head-slice experiments are retained only for reproduction analysis; the revised claims rely on audited reruns and explicit robustness checks over subset draws and redundancy layouts. Third, all final tables and figures are generated programmatically from raw CSV/JSON outputs rather than by manual summary packaging.

\subsection{Datasets, Models, and Seeds}

We use SST-2, MRPC, MNLI, and QQP from GLUE, together with AG News. The default setting uses TF--IDF features with logistic regression. We additionally report a small MLP and frozen DistilBERT feature settings in the extended experiments. For the strongest appendix-only feature setting, DistilBERT embeddings are extracted once and a linear classifier is trained on top.

End-to-end DistilBERT fine-tuning. For the transformer fine-tuning experiment, we fine-tune \texttt{distilbert-base-uncased} end-to-end on SST-2, MRPC, QQP, and MNLI subsets. The requested subset sizes are SST-2 train/validation $5000/1000$, MRPC $3500/600$, QQP $10000/1500$, and MNLI $10000/2000$; because the underlying validation splits are smaller for SST-2 and MRPC, the realized validation sizes are $872$ and $408$, respectively. Across all four tasks, we use 3 training epochs, batch size 16, learning rate $2\times 10^{-5}$, maximum sequence length 128, AdamW with weight decay 0.01, and a linear warmup-plus-decay schedule with warmup steps set to one tenth of the total training steps. This yields task-specific warmup counts of 93 for SST-2, 65 for MRPC, and 187 for both QQP and MNLI. We use three outer runs with seeds $\{11,23,37\}$, and each outer run forms a final aggregated ranking from five internal scoring seeds $\{42,123,456,789,1024\}$, i.e., aggregation budget $R=5$. Checkpoints are selected by best validation accuracy, and the reported per-example rankings are extracted from that best checkpoint using the loss, margin, and confidence proxy families. Here, the loss proxy is the negated per-example cross-entropy, the margin proxy is the gold-class probability minus the strongest competing class probability, and the confidence proxy is the gold-class probability. The purpose of this experiment is not to maximize task accuracy, but to test whether the ranking-stability phenomenon persists beyond TF--IDF/logistic-regression pipelines.

We distinguish between \emph{outer experimental runs} and \emph{internal aggregation seeds}. Unless otherwise noted, the main stability experiments use \(R=5\) aggregation seeds. For significance-critical comparisons, we additionally run 10 outer experimental runs and report paired confidence intervals and paired nonparametric tests over these outer runs.

\subsection{Redundancy Construction}

\paragraph{Exact duplicates.}
For exact-duplicate experiments, we sample source examples from the clean training set and append duplicate copies at a target redundancy rate \(r\). In the robustness experiments, both the subset draw and the redundancy layout are varied explicitly.

\paragraph{Synthetic near-duplicates.}
Synthetic near-duplicates follow the same source-sampling rule as exact duplicates, but each copied text is perturbed before re-featurization. The perturbation pipeline applies a stochastic combination of word dropout, local word swaps, word repetition, intra-sentence replacement, and segment shuffle. The revised paper uses \(r=0.3\) and \(s=0.6\) as the default audited operating point, and further reports sweeps over redundancy rate, perturbation strength, cluster size, cluster noise, and aggregation budget.

\paragraph{Natural redundancy.}
The main natural-redundancy showcase in the revised paper is QQP rather than the earlier weak SST-2 pair-only setting. We consider three cluster-discovery procedures: TF--IDF cosine thresholding, sentence-embedding similarity, and lexical character-\(n\)-gram similarity. The resulting clusters are approximate rather than oracle-quality, which makes this setting a more realistic test of redundancy-aware ranking.

\subsection{Proxies and SCARV Instantiations}

The base proxies considered in the revised paper are:
\begin{itemize}
    \item a \textbf{loss-based} proxy,
    \item a \textbf{TracIn-style} proxy,
    \item a \textbf{hybrid} proxy used in the method-decomposition study, and
    \item a \textbf{margin} proxy used for proxy-generality checks.
\end{itemize}

The default full \textsc{SCARV} instantiation follows the main-text method section: within-seed min--max normalization, a light diversity bonus, cluster-aware aggregation/allocation, and robust cross-seed aggregation via the median. To clarify the mechanism, we compare this full version against three nested alternatives:
\begin{itemize}
    \item \textbf{bare}: no structural step and no cross-seed aggregation,
    \item \textbf{seed-only}: robust aggregation across seeds without the structure-aware step, and
    \item \textbf{cluster-only}: the structure-aware step without cross-seed aggregation.
\end{itemize}

\subsection{Evaluation Metrics and Statistical Testing}
\label{app:statistics}

We report four families of metrics.

\paragraph{Alignment.}
We measure valuation alignment to leave-one-out (LOO) with Spearman correlation.

\paragraph{Global and local stability.}
Global ranking stability is computed differently for single-seed and aggregated methods. For single-seed bare rankings, global stability is computed from pairwise Spearman correlations across seed-specific rankings. For aggregation methods, each outer run first produces one final ranking from \(R\) internal scoring seeds, and stability is then computed across final rankings from different outer runs. Local stability is measured analogously with top-\(k\) and bottom-\(k\) Jaccard overlap.

\paragraph{Decision reproducibility.}
For subset selection, we report selected-subset overlap and selection gap in addition to downstream accuracy. For noisy-label retrieval, we report overlap of suspicious-example sets across runs.

\paragraph{Downstream utility.}
For subset selection, we report retained-subset test accuracy. For noisy-label retrieval, we report AUROC and, when available, AUPRC, Precision@\({k}\), Recall@\({k}\), and suspicious-set overlap.

All significance tests are paired over \emph{outer runs}. Pairwise seed-pair Spearman values are first summarized within each outer run and are \emph{not} treated as independent samples. For the key comparison between full \textsc{SCARV} and seed-median, we report paired bootstrap confidence intervals together with Wilcoxon signed-rank tests.

\section{Additional Results and Sensitivity Analyses}
\label{app:results}

This section reports appendix-level results from the \emph{audited rerun pipeline}. Our goal here is not to expand the paper's claims, but to stress-test the revised interpretation established in the main text. Across the analyses below, the same picture persists: stability-aware aggregation is robust to subset and layout variation, transfers beyond the original SST-2 configuration, and improves the reproducibility of ranking-based decisions. At the same time, the appendix sharpens the mechanism story: in the default synthetic regime, robust multi-seed aggregation is the dominant source of gain, while the structure-aware stage is most useful when redundancy is more informative, naturally mined, or sufficiently covered. The structure-quality diagnostics below further caution that stability gains alone do not validate the semantic correctness of a redundancy map.

\subsection{Alignment to Leave-One-Out}

The audited leave-one-out checks are diagnostic rather than central evidence. They support two limited conclusions. First, proxy choice matters: TracIn-style scores are generally more positively aligned with LOO than loss or margin scores, which are often weakly or negatively aligned. Second, stabilization and valuation alignment are not the same objective. Full \textsc{SCARV} substantially improves ranking stability, but its effect on LOO alignment is mixed across datasets and proxies. In some settings it preserves or slightly improves alignment; in others it reduces the LOO correlation relative to the bare proxy. We therefore interpret \textsc{SCARV} as a stability-oriented aggregation layer rather than as a replacement valuation functional.

...

These results matter for two reasons. First, they substantially weaken the concern that the paper is only a controlled SST-2 toy study. Second, they sharpen the regime story: while seed-only aggregation is strongest on average in the default synthetic decomposition, full \textsc{SCARV} becomes more competitive in realistic-task settings and under stronger feature representations.

\subsection{Mechanism Decomposition and Stress Tests}

The main text already reports the central mechanistic result: robust multi-seed aggregation is the primary source of stability gain in the default synthetic regime. Here we expand that comparison to include cluster-only variants, dedup-first baselines, and multiple synthetic stress regimes.

\begin{table*}[t]
    \centering
    \caption{\textbf{Method decomposition (Spearman stability, $\uparrow$).} All methods across six synthetic regimes and three proxies. Bold marks the best value per row. \emph{A.Dup} / \emph{O.Dup}: approximate / oracle dedup-then-rank. \emph{C.Only}, \emph{C.Div}, \emph{C.Col}, \emph{Unif}, \emph{Med}, \emph{Soft}: cluster-only variants. \emph{S.Mean}, \emph{S.Borda}, \emph{S.Med}: seed-only aggregators.}
    \label{tab:method_decomposition}
    \resizebox{\textwidth}{!}{%
    \small
    \setlength{\tabcolsep}{4pt}
    \begin{tabular}{llccccccccccccc}
        \toprule
        \textbf{Regime} & \textbf{Proxy} &
        \textbf{Bare} & \textbf{C.Only} & \textbf{C.Div} & \textbf{C.Col} & \textbf{Unif} & \textbf{Med} & \textbf{Soft} &
        \textbf{S.Mean} & \textbf{S.Borda} & \textbf{S.Med} &
        \textbf{Full SCARV} &
        \textbf{A.Dup} & \textbf{O.Dup} \\
        \midrule
        \multirow{3}{*}{Base}
          & Hybrid & 0.6845 & 0.6804 & 0.6296 & 0.6508 & 0.6804 & 0.6987 & 0.6820 & \textbf{0.9224} & 0.9207 & 0.8799 & 0.8502 & 0.6668 & 0.6508 \\
          & Loss   & 0.6040 & 0.6000 & 0.5393 & 0.6008 & 0.6000 & 0.6231 & 0.6018 & \textbf{0.8920} & 0.8880 & 0.8422 & 0.8018 & 0.5330 & 0.6008 \\
          & TracIn & 0.9564 & 0.9567 & 0.9584 & 0.9739 & 0.9567 & 0.9556 & 0.9566 & \textbf{0.9922} & 0.9914 & 0.9769 & 0.9774 & 0.9498 & 0.9739 \\
        \midrule
        \multirow{3}{*}{Noise-25}
          & Hybrid & 0.7412 & 0.7425 & 0.6729 & 0.6701 & 0.7425 & 0.7816 & 0.7424 & \textbf{0.9370} & 0.9344 & 0.8986 & 0.8627 & 0.6776 & 0.6701 \\
          & Loss   & 0.6724 & 0.6740 & 0.5892 & 0.5588 & 0.6740 & 0.7239 & 0.6748 & \textbf{0.9187} & 0.9138 & 0.8769 & 0.8339 & 0.5669 & 0.5588 \\
          & TracIn & 0.9642 & 0.9642 & 0.9665 & 0.9585 & 0.9642 & 0.9691 & 0.9642 & \textbf{0.9930} & 0.9923 & 0.9834 & 0.9851 & 0.9676 & 0.9585 \\
        \midrule
        \multirow{3}{*}{High Rate}
          & Hybrid & 0.7636 & 0.7598 & 0.6935 & 0.7039 & 0.7598 & 0.7773 & 0.7616 & \textbf{0.9444} & 0.9433 & 0.9117 & 0.8727 & 0.6556 & 0.7039 \\
          & Loss   & 0.7024 & 0.6969 & 0.6078 & 0.5517 & 0.6969 & 0.7259 & 0.6995 & \textbf{0.9266} & 0.9245 & 0.8913 & 0.8379 & 0.5866 & 0.5517 \\
          & TracIn & 0.9781 & 0.9785 & 0.9759 & 0.9663 & 0.9785 & 0.9775 & 0.9783 & \textbf{0.9940} & 0.9935 & 0.9868 & 0.9873 & 0.9762 & 0.9663 \\
        \midrule
        \multirow{3}{*}{Clust.\ Sz-4}
          & Hybrid & 0.7412 & 0.7391 & 0.6731 & 0.6723 & 0.7391 & 0.7508 & 0.7401 & \textbf{0.9370} & 0.9344 & 0.8986 & 0.8626 & 0.6776 & 0.6723 \\
          & Loss   & 0.6724 & 0.6691 & 0.5873 & 0.5570 & 0.6691 & 0.6888 & 0.6707 & \textbf{0.9187} & 0.9138 & 0.8769 & 0.8326 & 0.5669 & 0.5570 \\
          & TracIn & 0.9642 & 0.9644 & 0.9660 & 0.9512 & 0.9644 & 0.9645 & 0.9644 & \textbf{0.9930} & 0.9923 & 0.9834 & 0.9849 & 0.9676 & 0.9512 \\
        \midrule
        \multirow{3}{*}{Clust.\ Sz-8}
          & Hybrid & 0.7392 & 0.7370 & 0.6778 & 0.6810 & 0.7370 & 0.7436 & 0.7381 & \textbf{0.9316} & 0.9278 & 0.8924 & 0.8615 & 0.6960 & 0.6810 \\
          & Loss   & 0.6656 & 0.6621 & 0.5839 & 0.5635 & 0.6621 & 0.6720 & 0.6639 & \textbf{0.9130} & 0.9058 & 0.8665 & 0.8229 & 0.6136 & 0.5635 \\
          & TracIn & 0.9710 & 0.9710 & 0.9701 & 0.9673 & 0.9710 & 0.9709 & 0.9710 & \textbf{0.9916} & 0.9913 & 0.9799 & 0.9806 & 0.9749 & 0.9673 \\
        \midrule
        \multirow{3}{*}{Low Sim.}
          & Hybrid & 0.7279 & 0.7237 & 0.6554 & 0.6725 & 0.7237 & 0.7501 & 0.7261 & \textbf{0.9342} & 0.9315 & 0.8923 & 0.8564 & 0.6684 & 0.6725 \\
          & Loss   & 0.6702 & 0.6653 & 0.5897 & 0.5577 & 0.6653 & 0.6991 & 0.6678 & \textbf{0.9162} & 0.9114 & 0.8731 & 0.8300 & 0.5699 & 0.5577 \\
          & TracIn & 0.9654 & 0.9659 & 0.9660 & 0.9512 & 0.9659 & 0.9658 & 0.9656 & \textbf{0.9909} & 0.9901 & 0.9748 & 0.9775 & 0.9143 & 0.9512 \\
        \bottomrule
    \end{tabular}}
\end{table*}

The decomposition is informative in both positive and negative ways. Negatively, it rules out the claim that the structure-aware stage is the universally dominant component. Positively, it shows that full \textsc{SCARV} is still meaningfully above bare, cluster-only, and dedup-first baselines, and that structure becomes more useful when redundancy is more informative, naturally mined, or sufficiently covered. These results should not be read as evidence that the structure-aware stage dominates seed aggregation. Instead, they show that structural regularization remains meaningful relative to bare, cluster-only, and dedup-first alternatives, while repeated aggregation across seeds is the largest generic stabilizer.

\begin{figure}[tb]
    \centering
    \includegraphics[width=\columnwidth]{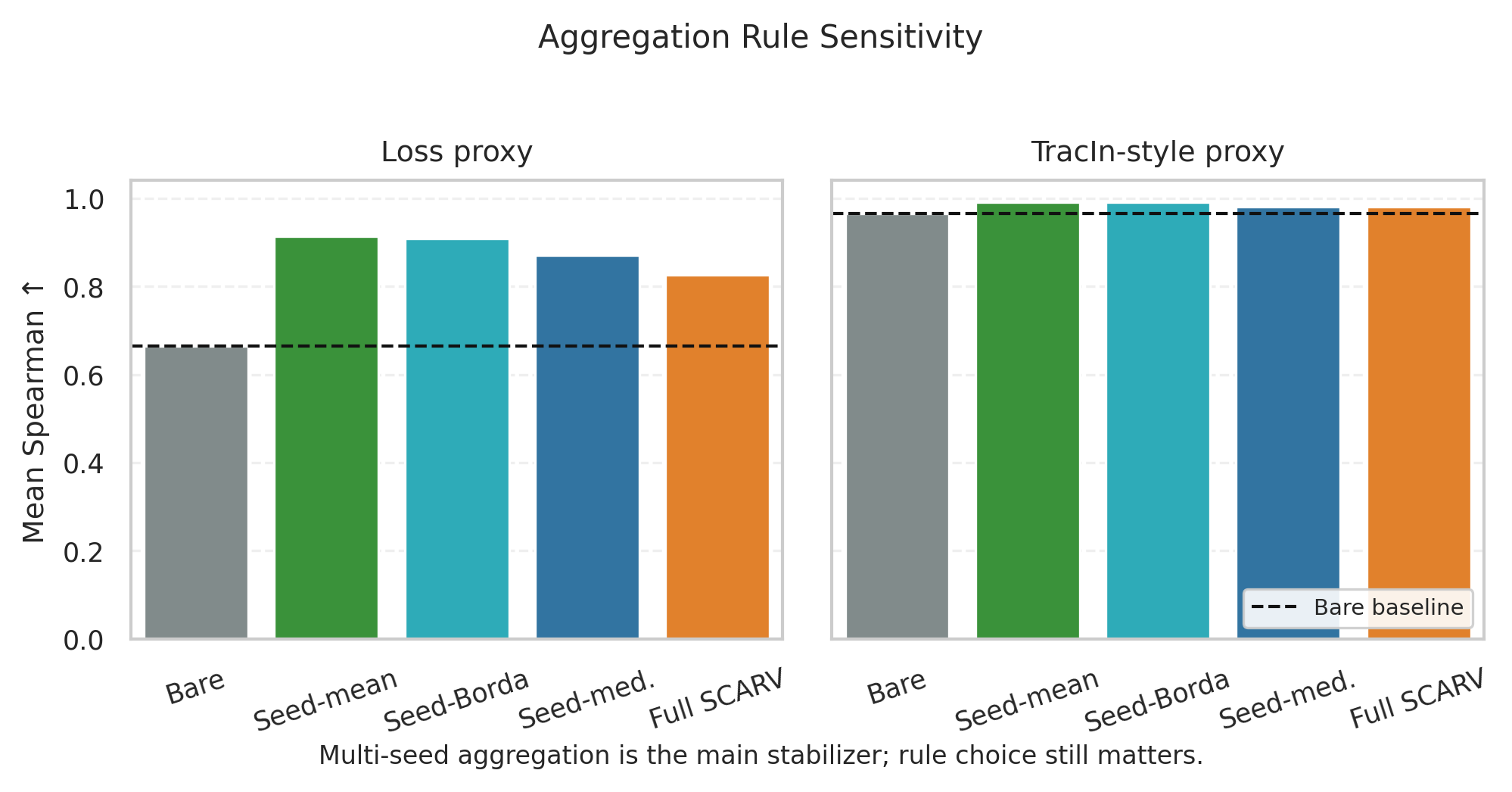}
    \caption{\textbf{Sensitivity to aggregation rule.} Adding multi-seed aggregation is the largest source of stability gain, while the choice among mean, median, and Borda still matters: mean and Borda are often stronger pure-stability baselines than the default median.}
\end{figure}

A major weakness of the original draft was that the natural-redundancy evidence on SST-2 was too weak to support strong claims. The revised appendix therefore uses QQP as the primary natural-redundancy showcase and compares three approximate cluster-discovery procedures.

\begin{table}[t]
    \centering
    \small
    \caption{\textbf{Natural redundancy on QQP.} Cluster statistics for three approximate discovery procedures.}
    \label{tab:app_natural_cluster_stats}
    \setlength{\tabcolsep}{3pt}
    \begin{tabular}{lcccccc}
        \toprule
        \textbf{Method} & \textbf{Cov.} & \textbf{\#Clust.} & \textbf{Mean} & \textbf{Max} & \textbf{Sim.} & \textbf{Pur.} \\
        \midrule
        TF--IDF cosine     & 0.026 & 36 & 2.17 & 5 & 0.910 & 0.972 \\
        Sent.\ embedding   & 0.065 & 74 & 2.65 & 8 & 0.921 & 0.942 \\
        Lexical char-$n$g  & 0.061 & 74 & 2.47 & 7 & 0.781 & 0.939 \\
        \bottomrule
    \end{tabular}
\end{table}

\begin{figure}[tb]
    \centering
    \includegraphics[width=\columnwidth]{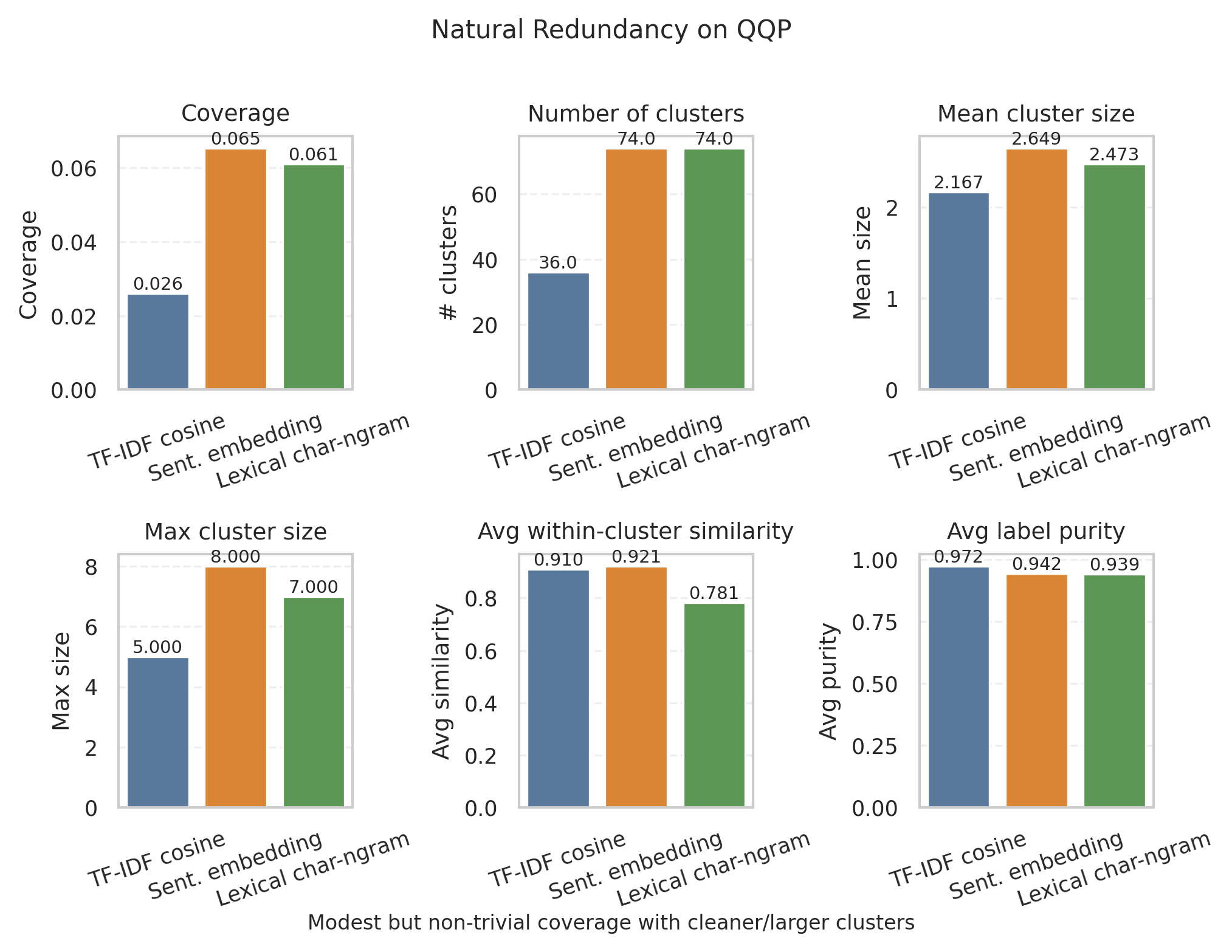}
    \caption{\textbf{Natural redundancy statistics on QQP.} The revised natural-redundancy setting has non-trivial coverage and substantially stronger redundancy structure than the earlier SST-2 pair-only regime.}
    \label{fig:app_natural_stats}
\end{figure}

The strengthened QQP results serve two purposes. First, they show that the paper is not restricted to synthetic duplicate injection. Second, they help identify when the structure-aware stage is most useful: redundancy-aware aggregation is most helpful when redundancy is more informative, naturally mined, or sufficiently covered. This is the strongest evidence in the paper that the structural stage can contribute value beyond generic cross-seed smoothing, even though best seed-only aggregation remains a strong upper baseline when enough repeated runs are available.

\subsection{Compute-Aware Seed-Budget Frontier}

The main text summarizes the compute-aware frontier; here we keep only a compact appendix summary rather than a separate giant table for each seed budget \(R\). For each budget, we compare full \textsc{SCARV} with the post-hoc strongest seed-only method under the same budget. The pattern is unchanged: full \textsc{SCARV} is often competitive at \(R=1\) or \(R=2\), but best seed-only aggregation dominates pure stability once \(R\) is moderately large.

\begin{table}[t]
    \centering
    \small
    \caption{\textbf{Compact compute-aware frontier summary.} $\Delta_{\mathrm{best}}$ is Full \textsc{SCARV} Spearman minus the post-hoc strongest seed-only method among seed-mean, seed-median, and seed-Borda under the same seed budget.}
    \label{tab:compute-frontier-appendix}
    \setlength{\tabcolsep}{3.5pt}
    \begin{tabular}{rrrrr}
        \toprule
        $R$ & \textbf{Settings} & \textbf{Full $>$ best upper} & \textbf{Mean $\Delta_{\mathrm{best}}$} & \textbf{Median $\Delta_{\mathrm{best}}$} \\
        \midrule
        1  & 18 & 16 & +0.0180 & +0.0233 \\
        2  & 18 & 15 & +0.0095 & +0.0126 \\
        3  & 18 & 0 & -0.0374 & -0.0335 \\
        5  & 18 & 0 & -0.0374 & -0.0307 \\
        7  & 18 & 0 & -0.0318 & -0.0251 \\
        10 & 18 & 0 & -0.0187 & -0.0138 \\
        \bottomrule
    \end{tabular}
\end{table}

\begin{figure}[tb]
    \centering
    \includegraphics[width=\columnwidth]{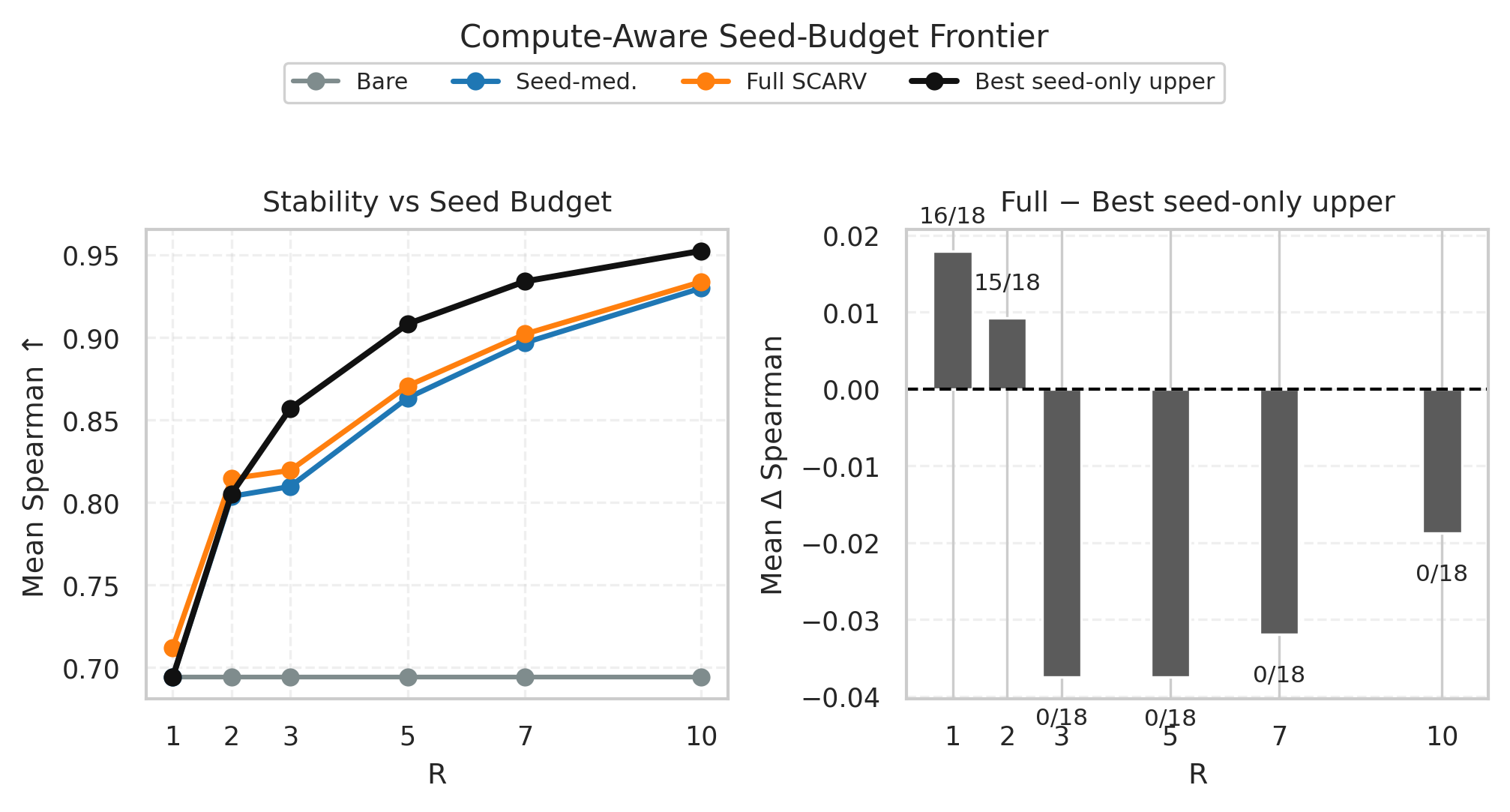}
    \caption{\textbf{Compute-aware seed-budget frontier.} Left: mean Spearman stability vs.~seed budget for bare, seed-median, Full \textsc{SCARV}, and the post-hoc best seed-only upper baseline. Right: mean $\Delta$ Spearman of Full \textsc{SCARV} relative to that best upper baseline; annotations show the number of settings in which Full \textsc{SCARV} exceeds it. Full \textsc{SCARV} is strongest mainly at very small budgets, while seed-only aggregation dominates once enough runs are available.}
    \label{fig:app_compute_frontier}
\end{figure}

\subsection{End-to-End Transformer Fine-Tuning}

To test whether the stability phenomenon is tied to shallow models, we additionally fine-tune \texttt{distilbert-base-uncased} end-to-end on SST-2, MRPC, QQP, and MNLI subsets. Across loss, confidence, and margin proxies, full \textsc{SCARV} consistently improves over bare rankings in Spearman stability, top-10 Jaccard overlap, and selected-subset reproducibility. The fine-tuning protocol and seed structure are described in Appendix~\ref{app:setup}. The results are also consistent with the main mechanism: full \textsc{SCARV} is competitive with seed-median, while seed-mean and seed-Borda remain stronger pure-stability baselines when repeated runs are available. Thus, the transformer experiment reduces the toy-setting concern without changing the paper's narrower interpretation.

\begin{table*}[p]
    \centering
    \small
    \caption{\textbf{End-to-end DistilBERT fine-tuning (MNLI).} Values are means over outer runs. We report Spearman stability, top-10 Jaccard overlap, and subset-overlap reproducibility at budget 0.3.}
    \label{tab:transformer_ft_mnli}
    \setlength{\tabcolsep}{4pt}
    \begin{tabular}{llccc}
        \toprule
        \textbf{Proxy} & \textbf{Method} & \textbf{Spearman} & \textbf{Top10 Jacc.} & \textbf{Subset overlap@0.3} \\
        \midrule
        Confidence & approx. dedup & 0.8707 & 0.5922 & 0.6712 \\
        Confidence & bare & 0.8732 & 0.5948 & 0.6759 \\
        Confidence & cluster-only & 0.8705 & 0.5956 & 0.6715 \\
        Confidence & full SCARV & 0.9356 & 0.6958 & 0.7662 \\
        Confidence & seed-Borda & 0.9700 & 0.7636 & 0.8246 \\
        Confidence & seed-mean & 0.9706 & 0.7674 & 0.8368 \\
        Confidence & seed-median & 0.9371 & 0.6860 & 0.7708 \\
        \midrule
        Loss & approx. dedup & 0.8707 & 0.5920 & 0.6712 \\
        Loss & bare & 0.8732 & 0.5948 & 0.6759 \\
        Loss & cluster-only & 0.8705 & 0.5959 & 0.6715 \\
        Loss & full SCARV & 0.9351 & 0.6952 & 0.7644 \\
        Loss & seed-Borda & 0.9700 & 0.7636 & 0.8246 \\
        Loss & seed-mean & 0.9708 & 0.7702 & 0.8388 \\
        Loss & seed-median & 0.9366 & 0.6854 & 0.7696 \\
        \midrule
        Margin & approx. dedup & 0.8682 & 0.5827 & 0.6681 \\
        Margin & bare & 0.8707 & 0.5839 & 0.6723 \\
        Margin & cluster-only & 0.8680 & 0.5853 & 0.6690 \\
        Margin & full SCARV & 0.9344 & 0.6837 & 0.7648 \\
        Margin & seed-Borda & 0.9694 & 0.7583 & 0.8226 \\
        Margin & seed-mean & 0.9699 & 0.7643 & 0.8365 \\
        Margin & seed-median & 0.9359 & 0.6833 & 0.7686 \\
        \bottomrule
    \end{tabular}
\end{table*}

\begin{table*}[p]
    \centering
    \small
    \caption{\textbf{End-to-end DistilBERT fine-tuning (MRPC).}}
    \label{tab:transformer_ft_mrpc}
    \setlength{\tabcolsep}{4pt}
    \begin{tabular}{llccc}
        \toprule
        \textbf{Proxy} & \textbf{Method} & \textbf{Spearman} & \textbf{Top10 Jacc.} & \textbf{Subset overlap@0.3} \\
        \midrule
        Confidence & approx. dedup & 0.9222 & 0.4575 & 0.7629 \\
        Confidence & bare & 0.9225 & 0.4463 & 0.7489 \\
        Confidence & cluster-only & 0.9213 & 0.4590 & 0.7590 \\
        Confidence & full SCARV & 0.9598 & 0.5545 & 0.8159 \\
        Confidence & seed-Borda & 0.9826 & 0.6490 & 0.8664 \\
        Confidence & seed-mean & 0.9834 & 0.7120 & 0.8935 \\
        Confidence & seed-median & 0.9597 & 0.5471 & 0.8044 \\
        \midrule
        Loss & approx. dedup & 0.9221 & 0.4575 & 0.7628 \\
        Loss & bare & 0.9225 & 0.4463 & 0.7489 \\
        Loss & cluster-only & 0.9209 & 0.4590 & 0.7591 \\
        Loss & full SCARV & 0.9589 & 0.5501 & 0.8139 \\
        Loss & seed-Borda & 0.9826 & 0.6490 & 0.8664 \\
        Loss & seed-mean & 0.9828 & 0.7176 & 0.8942 \\
        Loss & seed-median & 0.9589 & 0.5447 & 0.8026 \\
        \midrule
        Margin & approx. dedup & 0.9222 & 0.4575 & 0.7629 \\
        Margin & bare & 0.9225 & 0.4463 & 0.7489 \\
        Margin & cluster-only & 0.9213 & 0.4590 & 0.7590 \\
        Margin & full SCARV & 0.9598 & 0.5545 & 0.8159 \\
        Margin & seed-Borda & 0.9826 & 0.6490 & 0.8664 \\
        Margin & seed-mean & 0.9834 & 0.7120 & 0.8935 \\
        Margin & seed-median & 0.9597 & 0.5471 & 0.8044 \\
        \bottomrule
    \end{tabular}
\end{table*}

\begin{table*}[p]
    \centering
    \small
    \caption{\textbf{End-to-end DistilBERT fine-tuning (QQP).}}
    \label{tab:transformer_ft_qqp}
    \setlength{\tabcolsep}{4pt}
    \begin{tabular}{llccc}
        \toprule
        \textbf{Proxy} & \textbf{Method} & \textbf{Spearman} & \textbf{Top10 Jacc.} & \textbf{Subset overlap@0.3} \\
        \midrule
        Confidence & approx. dedup & 0.9241 & 0.6399 & 0.7982 \\
        Confidence & bare & 0.9242 & 0.6380 & 0.7973 \\
        Confidence & cluster-only & 0.9236 & 0.6398 & 0.7970 \\
        Confidence & full SCARV & 0.9591 & 0.7619 & 0.8621 \\
        Confidence & seed-Borda & 0.9807 & 0.8210 & 0.8987 \\
        Confidence & seed-mean & 0.9775 & 0.8143 & 0.8951 \\
        Confidence & seed-median & 0.9592 & 0.7607 & 0.8626 \\
        \midrule
        Loss & approx. dedup & 0.9246 & 0.6400 & 0.7983 \\
        Loss & bare & 0.9242 & 0.6380 & 0.7973 \\
        Loss & cluster-only & 0.9241 & 0.6399 & 0.7970 \\
        Loss & full SCARV & 0.9590 & 0.7625 & 0.8621 \\
        Loss & seed-Borda & 0.9807 & 0.8210 & 0.8987 \\
        Loss & seed-mean & 0.9763 & 0.8120 & 0.8942 \\
        Loss & seed-median & 0.9589 & 0.7613 & 0.8619 \\
        \midrule
        Margin & approx. dedup & 0.9241 & 0.6400 & 0.7982 \\
        Margin & bare & 0.9242 & 0.6380 & 0.7973 \\
        Margin & cluster-only & 0.9236 & 0.6399 & 0.7970 \\
        Margin & full SCARV & 0.9591 & 0.7619 & 0.8621 \\
        Margin & seed-Borda & 0.9807 & 0.8210 & 0.8987 \\
        Margin & seed-mean & 0.9775 & 0.8143 & 0.8951 \\
        Margin & seed-median & 0.9592 & 0.7607 & 0.8626 \\
        \bottomrule
    \end{tabular}
\end{table*}

\begin{table*}[p]
    \centering
    \small
    \caption{\textbf{End-to-end DistilBERT fine-tuning (SST-2).}}
    \label{tab:transformer_ft_sst2}
    \setlength{\tabcolsep}{4pt}
    \begin{tabular}{llccc}
        \toprule
        \textbf{Proxy} & \textbf{Method} & \textbf{Spearman} & \textbf{Top10 Jacc.} & \textbf{Subset overlap@0.3} \\
        \midrule
        Confidence & approx. dedup & 0.8494 & 0.5937 & 0.7311 \\
        Confidence & bare & 0.8529 & 0.5925 & 0.7356 \\
        Confidence & cluster-only & 0.8483 & 0.5987 & 0.7310 \\
        Confidence & full SCARV & 0.9184 & 0.7242 & 0.8201 \\
        Confidence & seed-Borda & 0.9649 & 0.7653 & 0.8576 \\
        Confidence & seed-mean & 0.9471 & 0.7840 & 0.8712 \\
        Confidence & seed-median & 0.9209 & 0.7232 & 0.8237 \\
        \midrule
        Loss & approx. dedup & 0.8494 & 0.5937 & 0.7311 \\
        Loss & bare & 0.8529 & 0.5925 & 0.7356 \\
        Loss & cluster-only & 0.8483 & 0.5987 & 0.7310 \\
        Loss & full SCARV & 0.9186 & 0.7282 & 0.8200 \\
        Loss & seed-Borda & 0.9649 & 0.7653 & 0.8576 \\
        Loss & seed-mean & 0.9439 & 0.7875 & 0.8670 \\
        Loss & seed-median & 0.9211 & 0.7236 & 0.8255 \\
        \midrule
        Margin & approx. dedup & 0.8494 & 0.5937 & 0.7311 \\
        Margin & bare & 0.8529 & 0.5923 & 0.7356 \\
        Margin & cluster-only & 0.8483 & 0.5985 & 0.7310 \\
        Margin & full SCARV & 0.9184 & 0.7242 & 0.8202 \\
        Margin & seed-Borda & 0.9649 & 0.7647 & 0.8576 \\
        Margin & seed-mean & 0.9471 & 0.7840 & 0.8712 \\
        Margin & seed-median & 0.9209 & 0.7232 & 0.8237 \\
        \bottomrule
    \end{tabular}
\end{table*}

\begin{figure}[tb]
    \centering
    \includegraphics[width=\columnwidth]{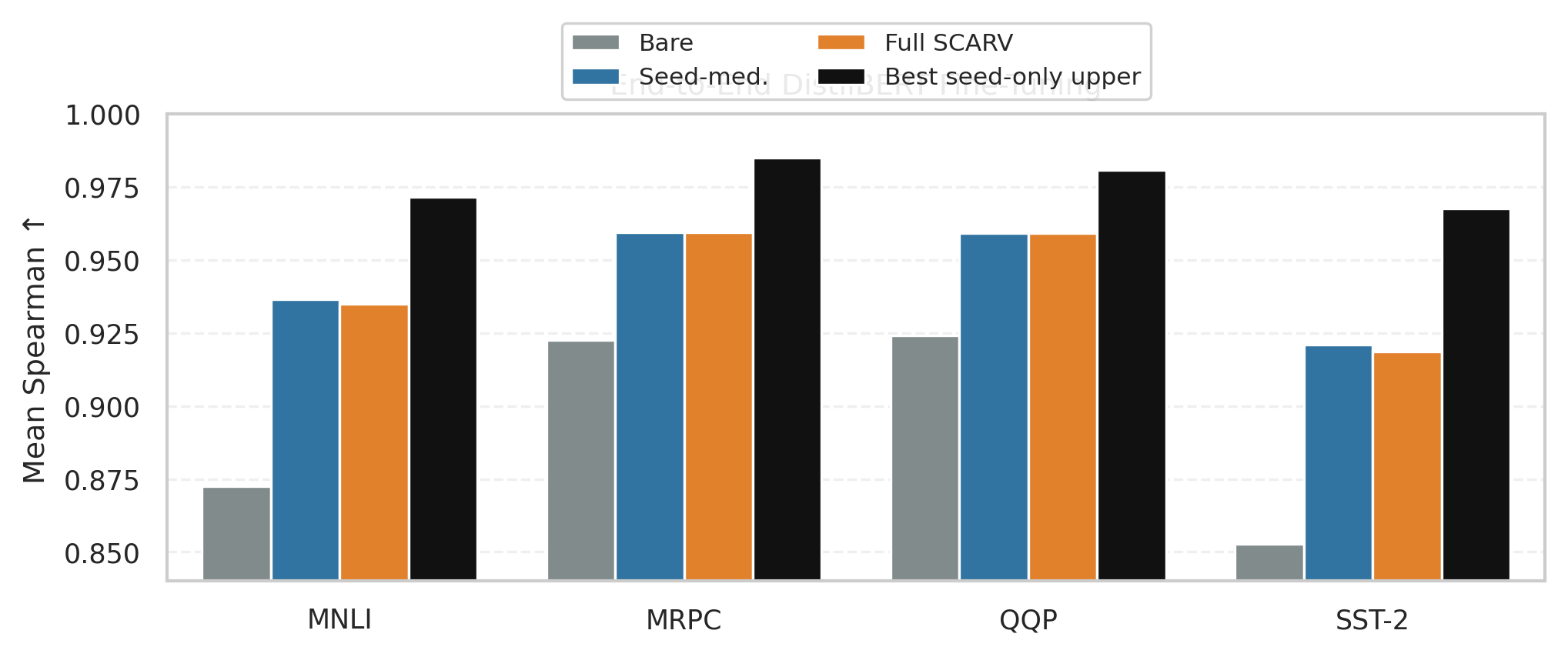}
    \caption{\textbf{End-to-end DistilBERT fine-tuning summary.} Mean Spearman stability across datasets for bare, seed-median, Full \textsc{SCARV}, and the post-hoc best seed-only upper baseline. Full \textsc{SCARV} substantially improves over bare and is competitive with seed-median, while the best seed-only upper baseline remains strongest for pure stability.}
    \label{fig:app_transformer_summary}
\end{figure}

\subsection{Practical Value and Proxy Generality}

The main-text practical-value result is that stabilization helps most through \emph{decision reproducibility}. Here we give the full appendix-level numbers for selection variance and noisy-label retrieval.

\begin{table}[t]
    \centering
    \small
    \caption{\textbf{Noisy-label retrieval (AUROC, $\uparrow$) on SST-2.} Results for loss and TracIn proxies at noise rates 10\% and 20\%. Bold marks the highest value per row.}
    \label{tab:selection_noise_retrieval}
    \setlength{\tabcolsep}{5pt}
    \begin{tabular}{llcccc}
        \toprule
        \textbf{Proxy} & \textbf{Noise} & \textbf{Bare} & \textbf{Seed-median} & \textbf{Full SCARV} \\
        \midrule
        Loss   & 10\% & 0.7044 & \textbf{0.7442} & 0.7115 \\
        Loss   & 20\% & 0.6563 & \textbf{0.6946} & 0.6724 \\
        TracIn & 10\% & \textbf{0.5863} & 0.5859 & 0.5601 \\
        TracIn & 20\% & 0.6081 & \textbf{0.6117} & 0.5872 \\
        \bottomrule
    \end{tabular}
\end{table}

Across both subset selection and noisy-label retrieval, the most consistent benefit is that ranking-based decisions become more repeatable. Selected subsets overlap more across runs, suspicious-example sets overlap more across runs, and selection gaps shrink, even when mean downstream accuracy or AUROC changes only modestly. This is why the paper frames practical value through reproducibility rather than through a claim of uniform utility gains.

We also test whether the effect generalizes to an additional proxy family.

\begin{table}[t]
    \centering
    \small
    \caption{\textbf{Proxy generality (SST-2, Spearman, $\uparrow$).} Stability results across loss, TracIn, and margin proxies. Bold marks the highest value per row.}
    \label{tab:proxy_generality}
    \setlength{\tabcolsep}{5pt}
    \begin{tabular}{lccc}
        \toprule
        \textbf{Proxy} & \textbf{Bare} & \textbf{Seed-median} & \textbf{Full SCARV} \\
        \midrule
        Loss   & 0.6040 & \textbf{0.8422} & 0.8018 \\
        Margin & 0.5413 & \textbf{0.8028} & 0.7791 \\
        TracIn & 0.9564 & 0.9769          & \textbf{0.9774} \\
        \bottomrule
    \end{tabular}
\end{table}

The margin-proxy results support the same qualitative conclusion as the loss and TracIn families: the aggregation effect is real, but the strongest gains on noisier proxies come from repeated robust aggregation across seeds.

\subsection{Statistical Reliability}

Finally, we report paired statistical tests for the key comparison between full \textsc{SCARV} and seed-median.

\begin{table}[t]
    \centering
    \small
    \caption{\textbf{Paired significance: Full \textsc{SCARV} vs.\ Seed-median.} Mean $\Delta$ Spearman, 95\% bootstrap CI, and Wilcoxon signed-rank $p$-value over outer configurations. Significant results ($p < 0.05$) marked with $^*$.}
    \label{tab:significance_key}
    \setlength{\tabcolsep}{4pt}
    \begin{tabular}{lcccc}
        \toprule
        \textbf{Dataset} & \textbf{Mean $\Delta$} & \textbf{95\% CI} & \textbf{$p$-value} \\
        \midrule
        SST-2   & $-0.0121$ & $[-0.0178,\ -0.0060]$ & $0.0004^*$ \\
        AG News & $-0.0011$ & $[-0.0092,\ +0.0070]$ & $0.5609$ \\
        MRPC    & $+0.0098$ & $[+0.0054,\ +0.0142]$ & $0.1250$ \\
        MNLI    & $+0.0059$ & $[+0.0049,\ +0.0069]$ & $0.1250$ \\
        QQP     & $+0.0140$ & $[+0.0127,\ +0.0154]$ & $0.0312^*$ \\
        \bottomrule
    \end{tabular}
\end{table}

These tests compare full \textsc{SCARV} to seed-median, not to the post-hoc best seed-only upper baseline. The statistical results support a deliberately nuanced claim. On the audited SST-2 synthetic regime, seed-median is significantly stronger than full \textsc{SCARV}. On QQP natural redundancy, the sign reverses in favor of full \textsc{SCARV}. On MRPC and MNLI, full \textsc{SCARV} is slightly ahead but not significant under the current sample size. Thus, the revised appendix supports a regime-dependent interpretation rather than a universal ranking of methods.

\section{Extended Related Work and Additional Discussion}
\label{app:related}

\subsection{Extended Related Work}

\paragraph{Additional value notions and structured valuation settings.}
Beyond the valuation literature discussed in the main paper, several works broaden the design space of what it means to assign value to training examples. These include task-specific valuation for nearest-neighbor models and subsequent weighted or thresholded neighborhood-based extensions \citep{jia2019efficient_vldb,wang2024weightedknn,wang2023thresholdknn}, class-wise value decompositions such as CS-Shapley \citep{schoch2022csshapley}, and Shapley-style reasoning in adjacent uncertainty settings such as conformal prediction \citep{jaramillo2021shapleyconformal}. These extensions are relevant because they show that ``value'' itself is not monolithic. Our paper is orthogonal to that axis: \textsc{SCARV} does not redefine the value notion, but stabilizes the ranking induced by a chosen proxy when redundancy and stochastic training make pointwise scores unreliable.

\paragraph{Attribution beyond the main-text baselines.}
The main paper focuses on loss-based and TracIn-style proxies, but the broader attribution literature now includes scalable or generalized formulations such as DataInf, approximate-unrolling-based attribution, multiply-robust causal change attribution, and generalized influence-style methods for non-decomposable losses \citep{kwon2024datainf,bae2024approxunroll,quintas2024multiply,deng2025versatile}. These methods make example-level attribution increasingly practical in larger or more structured settings. Their relevance to \textsc{SCARV} is again complementary rather than competitive: they provide richer proxy mechanisms, while \textsc{SCARV} addresses the separate question of how those proxy-induced rankings should be aggregated under repeated or near-repeated structure.

\paragraph{Diagnostics, memorization, and repeated structure.}
A further neighboring literature studies ranking-like signals for dataset diagnostics, memorization, and instability. In addition to example forgetting and dataset cartography, zero-shot data maps and recent memorization-tracking work show that example-level behavior can reflect ambiguity, noise, recency, and repeated exposure in ways that are consequential but not reducible to a single scalar notion of value \citep{basile2023zeroshot,jagielski2023measuring,leybzon2024learning}. At the corpus level, repeated content also interacts with privacy and copyright concerns in large language models \citep{meeus2024copyright}. This broader view supports the motivation for our paper: under redundancy, reproducibility of a proxy-induced ranking is itself a meaningful property, even when the downstream use of the ranking varies across attribution, debugging, pruning, or curation.

\subsection{Additional Discussion on Proxy--Objective Mismatch}

One of the clearest empirical patterns in the paper is that influence-oriented and utility-oriented proxies can disagree sharply. In our diagnostic checks, TracIn-style scores are generally more positively aligned with LOO-style valuation than loss or margin scores, but the effect of stabilization on LOO alignment is mixed. TracIn-style scores are also often weak direct selectors under aggressive pruning. Loss-based scores exhibit the opposite pattern more often: they can be weaker valuation signals while remaining competitive or stronger for subset utility under some budgets. This mismatch should not be interpreted as a failure of the framework. It reflects a genuine difference between what different proxy families are designed to measure.

From this perspective, \textsc{SCARV} does not attempt to collapse all ranking signals into one universal ordering. Rather, it stabilizes the behavior of a chosen proxy under redundancy and stochastic training. If the underlying proxy is attribution-oriented, \textsc{SCARV} makes that attribution-oriented ranking more reproducible. If the underlying proxy is utility-oriented, \textsc{SCARV} makes that utility-oriented ranking more reproducible. This reinforces the paper's main separation between valuation alignment and ranking stability.

\subsection{Additional Discussion on Mechanism and Scope}

The appendix-level decomposition results materially sharpen how the method should be interpreted. In the default synthetic regime, the strongest methods are seed-only aggregators such as seed-mean, seed-Borda, and seed-median. This makes the contribution narrower than a claim of universal dominance, but stronger as an empirical statement about what actually causes ranking instability under redundancy. The main source of gain is repeated robust aggregation across seeds; the structure-aware stage contributes when redundancy is more informative, naturally mined, or sufficiently covered, especially at small aggregation budgets. The compute-aware frontier further shows that this contribution is most visible at small aggregation budgets; when many repeated proxy runs are available, seed-mean or seed-Borda usually becomes the strongest pure-stability baseline. The structure-quality diagnostics add a further caveat: high stability under a structural regularizer does not by itself certify that the clusters are semantically meaningful.

We view this as a useful design lesson for data-centric NLP. Rather than expecting one method to define value, stabilize rankings, and optimize every downstream objective simultaneously, a more realistic pipeline is modular: choose a proxy for a purpose, then stabilize it under redundancy. The revised results support that more modest but more defensible picture.

\subsection{Limitations and Future Directions}

Even after the revised experiments, several limitations remain. First, the strongest support in the paper concerns ranking stability and decision reproducibility, not universal gains in downstream selection utility. Second, although the natural-redundancy evidence is now much stronger than in the original draft, the paper still stops well short of large-scale end-to-end pretraining or instruction-tuning workflows. Third, the current framework assumes that some usable redundancy structure can be injected, mined, or approximately recovered; in many realistic settings, redundancy is softer, noisier, or hierarchical. The structure-aware component also depends on the quality and interpretation of the redundancy map. Our structure-quality diagnostics show that pure ranking stability can remain high even under random or degraded cluster assignments, so stability alone should not be taken as evidence that the clusters are semantically meaningful. \textsc{SCARV} should therefore be used with cluster-quality auditing rather than arbitrary partitions. The compute frontier also relies on a post-hoc best seed-only upper baseline. This is useful for stress-testing \textsc{SCARV} against the strongest pure-stability seed aggregator, but future work should study how practitioners should choose aggregation rules without access to post-hoc stability labels.

These limitations point to concrete future directions. One is to move from hard clusters to soft or uncertain redundancy maps. A second is to study stability-oriented aggregation in larger end-to-end fine-tuning pipelines. A third is to design structure-aware aggregators that adapt more explicitly to cluster size, cluster confidence, or proxy family. More broadly, the revised evidence suggests that stability under redundancy is a distinct dimension of data-centric method design, rather than a side effect that can be left implicit.

\end{document}